\documentclass[twocolumn,trackchanges]{aastex631}
\usepackage{multirow}

\received{September 10, 2024}
\revised{November 12, 2024}
\accepted{November 14, 2024}

\shorttitle{SED Modeling of BL Lacertae During Submm Outburst and Low X-Ray Polarization State}
\shortauthors{Mondal et al.}

\begin{document}

\title{Spectral Energy Distribution Modeling of BL Lacertae During a Large Submillimeter Outburst and Low X-Ray Polarization State}

\author[0009-0007-8215-6031]{Ayon Mondal} 
\affiliation{School of Astrophysics, Presidency University, 86/1 College Street, Kolkata 700073, India.}
\email{E-mail: mondalayon2001@gmail.com}
\author[0009-0005-6387-4789]{Arijit Sar}
\affiliation{School of Astrophysics, Presidency University, 86/1 College Street, Kolkata 700073, India.}
\author[0009-0007-7463-5147]{Maitreya Kundu}
\affiliation{School of Astrophysics, Presidency University, 86/1 College Street, Kolkata 700073, India.}
\affiliation{Department of Physics, Washington University in St.\ Louis, 1 Brookings Drive, St.\ Louis, MO 63130, USA.} 
\author[0000-0001-9899-7686]{Ritaban Chatterjee}
\affiliation{School of Astrophysics, Presidency University, 86/1 College Street, Kolkata 700073, India.}
\author[0000-0002-5481-5040]{Pratik Majumdar}
\affiliation{Saha Institute of Nuclear Physics, A CI of Homi Bhabha National Institute, Kolkata 700064, West Bengal, India.}

\begin{abstract}

In 2023 October-November, the blazar BL Lacertae underwent a very large-amplitude submm outburst. The usual single-zone leptonic model with the lower energy peak of the spectral energy distribution (SED) fit by the synchrotron emission from one distribution of relativistic electrons in the jet and inverse-Compton (IC) scattering of lower energy photons from the synchrotron radiation in the jet itself (synchrotron self-Compton or SSC) or those from the broad line region and torus by the same distribution of electrons cannot satisfactorily fit the broadband SED with simultaneous data at submm--optical--X-ray--GeV energies. Furthermore, simultaneous observations with IXPE indicate the X-ray polarization is undetected. We consider two different synchrotron components, one for the high flux in the submm wavelengths and another for the data at the optical band, which are supposedly due to two separate distributions of electrons. In that case, the optical emission is dominated by the synchrotron radiation from one electron distribution while the X-rays are mostly due to SSC process by another, which may result in low polarization fraction due to the IC scattering. We show that such a model can fit the broadband SED satisfactorily as well as explain the low polarization fraction at the X-rays.

\end{abstract}

\keywords{galaxies: active --- galaxies: jets --- quasars: general --- blazars --- individual: BL Lacertae}

\section{Introduction} \label{sec:intro}

One of the main characteristics of radio-loud active galactic nuclei (AGNs) is their narrow, bipolar relativistic jets, which contain highly energetic particles originating close to the central engine and moving approximately perpendicular to the accretion flow. Blazars are a subclass of radio-loud AGN in which the jet axis is pointed towards the observer at a very small angle ($\lesssim 10^{\circ}$) to the line of sight \citep{Urry_1982, Urry_1995} causing the jet to be strongly beamed and the counter-jet correspondingly de-beamed in the observer's frame. This makes blazars detectable up to high redshifts \citep[see, \textit{e.g.},][]{Ghisellini_2011, Ackermann_2017, Sahakyan_2020, Banados_2024}. Blazar emission is thus dominated by the jets, which emit non-thermal continuum emission across a large range of wavelengths, sometimes from radio to TeV $\gamma$-rays and is variable at a range of timescales from hours to decades. The double-humped SED \citep{Fossati_1998} of a blazar consists of a low energy emission component (radio to UV and sometimes extending to X-rays) produced by the relativistic electrons present in the jet via the synchrotron process \citep{Blandford_1978, Kirk_1998, Potter_2012, Bloom_Marscher_1996}. In the so-called `leptonic scenario', the higher energy hump,  which peaks at GeV-TeV energies, may be modeled with inverse-Compton (IC) scattering by the same electrons. The photons produced by the synchrotron emission in the jet itself may get up-scattered by the energetic electrons present in the jet to higher energies by the IC mechanism, called synchrotron self-Compton \citep[SSC;][]{Ghisellini_1985, Maraschi_1992, Bloom_Marscher_1996, Chiang_2002}. Low energetic external photons from the broad line region (BLR), torus, and sometimes from the accretion disk, can also get up-scattered by those electrons, to higher energies, by the external Compton \citep[EC;][]{Sikora_1994, Blazejowski_2000, Dermer_2009, Ghisellini_2009} process. In the context of the `hadronic scenario', protons in the jet may be accelerated to ultra-relativistic energies and can contribute to the emission of higher energy radiation through various mechanisms including synchrotron processes, proton-initiated cascades, and interactions of secondary particles with photons \citep{mucke01, Bottcher_2004, bottcher13, botta16, Ackermann_16, Abdo_2011_mrk421}.

UV radiation from the accretion disk photoionizes the BLR clouds and causes the broad emission lines seen in the optical spectrum \citep{Kwan_1981}. Depending upon the emission lines in the optical spectra of blazars, they can be classified into two sub-classes \citep{Urry_1995}: BL Lacertae objects (BL Lacs), which exhibit a featureless optical spectrum with very weak emission lines (equivalent width, EW $\leq 5 \text{\AA}$) and flat-spectrum radio quasars (FSRQs), in which the emission lines are stronger. On the other hand, depending upon the synchrotron peak frequency ($\nu_{\text{sync}}$), blazars are classified into three categories: the synchrotron hump peaks at UV--X-ray ($\nu_{\text{sync}} > 10^{15} \text{Hz}$) regime and beyond in high-synchrotron peaked (HSP) blazars, at infrared energies ($\nu_{\text{sync}} < 10^{14} \text{Hz}$) in low-synchrotron peaked (LSP) blazars, and in between for the intermediate-synchrotron peaked (ISP) blazars \citep{Abdo_2010}. Usually, most of the LSP blazars are FSRQ type and are more luminous than BL Lacs. 

BL Lacertae \citep[Fermi catalog name: 4FGL J2202.7+4216;][]{Ajello_2020}, the eponymous blazar of the BL Lac class, situated at a redshift of $z = 0.0668 \pm 0.0002$ \citep{Hawley_1977} is a typically bright ISP blazar that shows rapid variability in a wide energy range including optical and radio frequencies \citep{Marscher_2008_Nature,Larionov_2010,Wehrle_2016}. It is commonly classified as a BL Lac type blazar with low to intermediate synchrotron peak frequency \citep{Nilsson_2018, Ackermann_2011} although an HSP-type behavior has been observed in this source during some epcohs, e.g., 2020 October \citep{Sahakyan_2022}. 

Over the past decade, BL Lacertae has been regularly monitored in the VHE $\gamma$-rays, X-ray, UV, optical, and radio wavelengths by several space and ground-based observing programs. Therefore, well-sampled multi-epoch SEDs have been constructed, which have been modeled successfully by several authors \citep{Raiteri_2009,Raiteri_2013,MAGIC_2019,Weaver_2020}, which resulted in a better understanding of the emission properties of the source in different bands over a longer time period. 
In those models, the lower energy hump is generated by synchrotron radiation and the higher energy part of the SED is modeled by the SSC process only, without any EC component in the IC scattering. The absence of any EC contribution is justified by the lack of significant emission from the BLR or torus possibly due to a faint accretion disk. However, the minimal presence of emission lines in the observed optical spectrum of BL Lac may be attributed to the beaming effect, which enhances the jet radiation in the direction of the observer and hence emission lines are overwhelmed by the beamed continuum. Presence of weak optical lines (H{\small $\alpha$} and H{\small $\beta$}) in the spectrum of BL Lac \citep{Corbett_1996, Capetti_2010} indicates the presence of a BLR, and can lead to FSRQ-like behaviors. In some epochs, particularly when VHE $\gamma$-ray emission was detected from BL Lac, SED is better fit by adding an EC component along with the SSC component \citep{Bloom_1997,Madejski_1999,Bottcher_2000,Abdo_2011,MAGIC_2019}. 


In 2023 November, BL Lac underwent a very large outburst at the submm wavelength \citep{2023ATel16340....1G} with flux density reaching 21 Jy, more than 30\% higher than any previous flare by the source in the 20-year-long light curve measured by the the Submillimeter Array (SMA) and it also brightened significantly at multiple other wave bands. While high-amplitude variability at weeks-months timescale is not uncommon in numerous blazars, this event was unique because the X-ray polarization properties of the source could be observed simultaneously using the Imaging X-Ray Polarimetry Explorer (IXPE). In this manuscript, we construct a well-sampled broadband SED of BL Lac during that unprecedented submm outburst using data from simultaneous observations of \textit{Fermi}-LAT, \textit{Swift} X-ray Telescope (XRT), \textit{Swift} Ultraviolet and Optical Telescope (UVOT), IXPE, Nuclear Spectroscopic Telescope Array (NuSTAR), Submillimiter Array (SMA), Perkins Telescope Observatory (PTO) and Very Long Baseline Array (VLBA). We model the SED with nonthermal emission generated by energetic electrons present in the jet of BL Lac and investigate whether the physical scenario implied by our best-fit SED model is consistent with the X-ray polarization observations.

\begin{table*}
\centering
\caption{Multiwavelength simultaneous observations of BL Lacertae during 2023 November presented in the SED}
\label{tab:observation}
\begin{tabular}{c | c | c}
\hline
\textbf{Facilities} & \textbf{Observation Date (MJD)} & \textbf{Band / Energy range}  \\
\hline
\hline
\textit{Fermi}-LAT & $60218 - 60279$ & $0.1 - 300$ GeV  \\
\textit{NuSTAR} & 60261 & $0.3 - 78$ keV  \\
IXPE & $60255 - 60265$ & $2 - 8$ keV  \\
\textit{Swift}-XRT & $60253 - 60276$ & $0.3 - 10$ keV  \\
\textit{Swift}/UVOT & $60260 - 60262$ & UVW2, UVM2, UVW1, and U, B, and V \\
Perkins & 60261 & B, V, I, and R  \\
SMA & 60254, 60257, 60258, 60262, 60265, 60271 & $1.4 - 1.1$ mm  \\
VLBA & 60254, 60273 & 43 GHz  \\
VLBA & 60262 & 15 GHz  \\
\hline
\end{tabular}
\end{table*}

This paper is organized as follows. Observation and data reduction methods are described in \S \ref{sec:data_reduc}. The broadband SED modeling and its results are described in \S \ref{sec:SED_modeling}. We discuss the results in \S \ref{sec:Discussion}.

\section{Observation, Data Reduction and Analysis} \label{sec:data_reduc}

In this section, we present simultaneous multi-wavelength observations of BL Lacertae in 2023 November, from radio to $\gamma$-ray energies. See Table \ref{tab:observation} for a summary of the data. Figure \ref{fig:LCs} presents the multi-wavelength light curves of BL Lacertae from observations made by \textit{Fermi}-LAT, \textit{Swift}-XRT, SMA, and VLBA at 15 GHz. Below, we discuss the reduction and analyses of data from different facilities.
\subsection{$\gamma$-ray Data}

\begin{figure}
\centering
\includegraphics[width=\columnwidth]{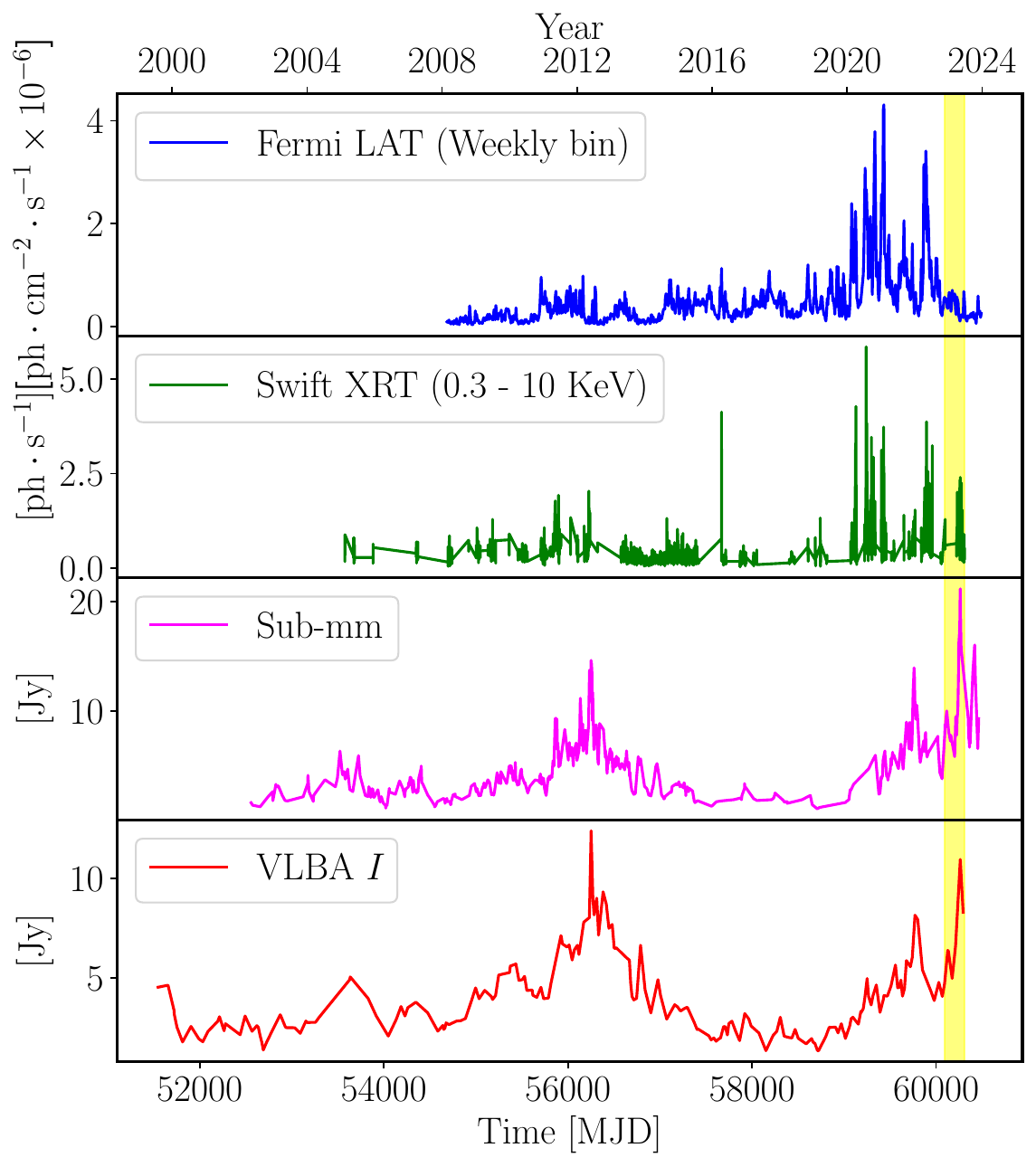}
\caption{Multiwavelength light curve of BL Lacertae observed by Fermi-LAT (weekly binning), Swift XRT ($0.3 - 10$ KeV), Submm ($1.4 - 1.1$ mm), and VLBA (15 GHz). The yellow strip is the epoch of our interest during the historical highest submm flare (2023 October-November).}
\label{fig:LCs}
\end{figure}

In this study, the publicly available $\gamma$-ray data between MJD 60218-60279 are accumulated from the Fermi-LAT data server\footnote{\url{https://fermi.gsfc.nasa.gov/ssc/data/access/}} and analyzed using the Fermi Science Tools \citep[version 2.2.0;][]{Fermitools}. We performed an unbinned likelihood analysis using the Pass8 `Source’ class events with a higher probability of being photons (\texttt{evclass} = 128, \texttt{evtype} = 3) in the energy range from 100 MeV to 300 GeV using the
\texttt{P8R3\_SOURCE\_V3} instrument response function. The events were selected from a circular region of interest of radius 12$^\circ$ around the $\gamma$-ray position of BL Lac. In the model, we used \textit{Fermi}-LAT fourth source catalog Data Release 3 \citep[4FGL-DR3;][]{Abdollahi2022} and around the target Galactic (\texttt{gll\_iem\_v07)} as well as the isotropic (\texttt{iso\_P8R3\_SOURCE\_V3\_v1}) diffuse emission components were considered, and the source spectrum was modeled using a log-parabola function. The weekly binned \textit{Fermi}-LAT light curve shown in Figure \ref{fig:LCs} are obtained from the publicly available \textit{Fermi}-LAT light curve repository\footnote{\url{https://fermi.gsfc.nasa.gov/ssc/data/access/lat/LightCurveRepository/index.html}}.

\subsection{X-ray Data}
\subsubsection{NuSTAR}

\begin{figure*}
    \centering
     \includegraphics[width=0.49\textwidth]{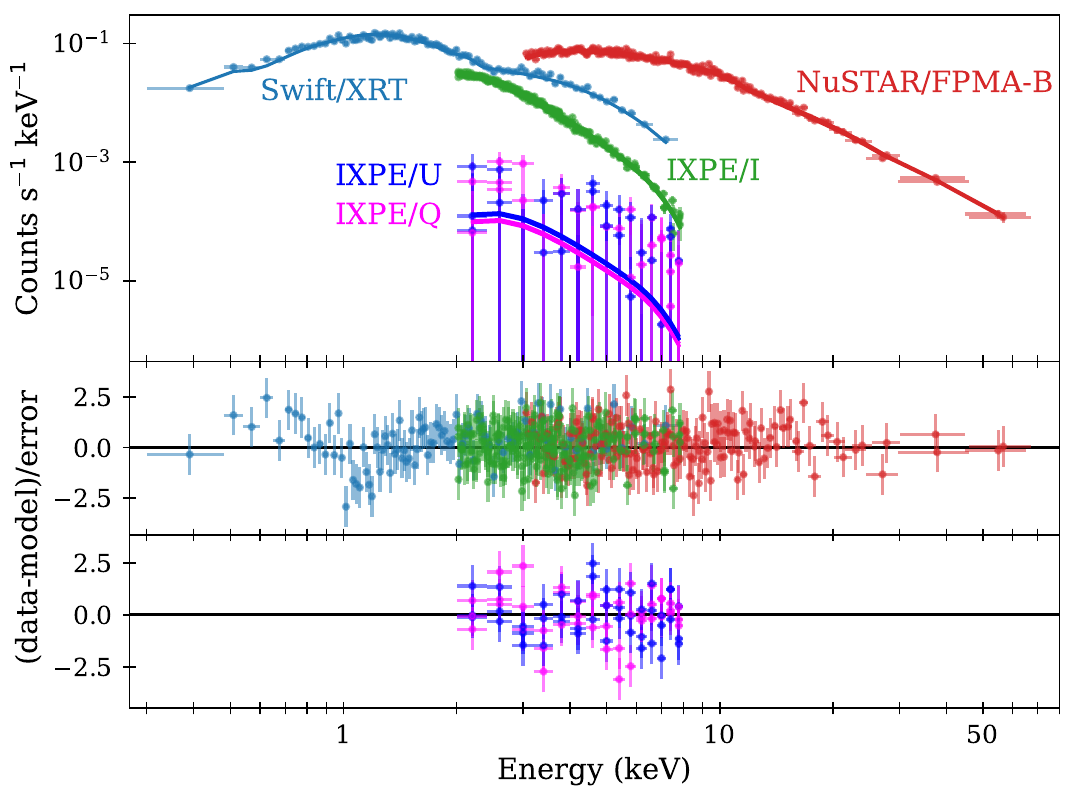}
    \includegraphics[width=0.49\textwidth]{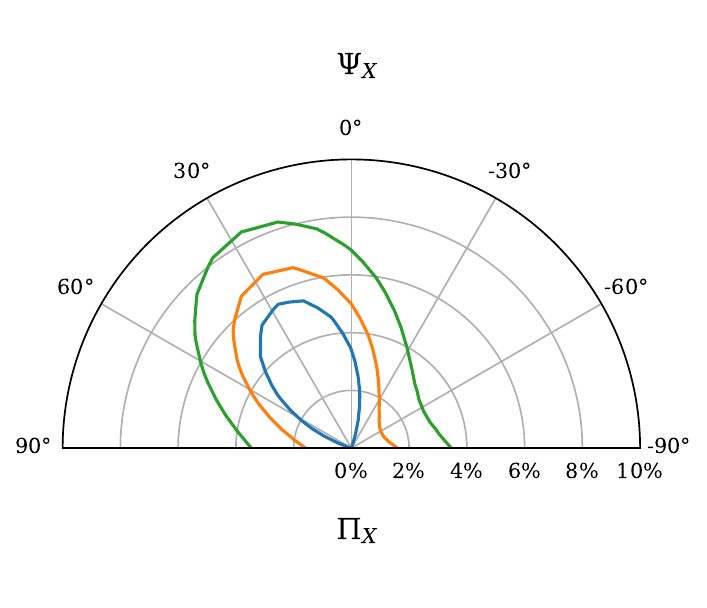}
    \caption{Left panel: Spectro-polarimetric joint fitting of X-ray data obtained from IXPE, \textit{Swift}-XRT, and NuSTAR using XSPEC with a power-law model. Right panel: Confidence contours 68\% (blue), 90\% (orange), and 99\% (green) of parameters $\psi_X$ vs $\Pi_X$ obtained from the spectro-polarimetric fit.}
    \label{fig: X-ray pol}
\end{figure*}

BL Lac was observed with \textit{NuSTAR} high-energy X-Ray mission \citep{Harrison_NuSTAR} on 2023 November 13 (Obs. Id. 80901639002; exposure time = 20.86 ks) during the submm flaring activity. The \textit{NuSTAR} data obtained from HEASARC archive\footnote{\url{https://heasarc.gsfc.nasa.gov/docs/nustar/nustar_archive.html}} was processed with the NuSTAR Data Analysis Software  consolidated in the HEASoft (v.6.32.1) software package. Source events were extracted using a circular region of size $70^{\prime \prime}$ and background events were extracted from a source-free region of size $90^{\prime \prime}$. We extracted the spectra for both Focal Plane Module A (FPMA) and B (FPMB) instruments onboard \textit{NuSTAR}. To ensure Gaussian statistics, spectra from both modules were grouped using \texttt{grppha} to ensure at least 50 counts per spectral bin. During spectral fitting we considered the  X-ray spectra from both modules in the energy range 3$-$78 keV.

\subsubsection{Swift-XRT}

BL Lac was observed throughout 2023 November by \textit{Swift}. To increase the signal-to-noise ratio, we extracted X-ray spectrum of BL Lacertae from the combined data of all photon counting (PC) mode observations (exposure time: 45.7 ks) performed in 2023 November using the \textit{Swift}-XRT \citep{Swift_XRT}. The combined X-ray spectrum was created using the online \textit{Swift}-XRT data product generator tool\footnote{\url{https://www.swift.ac.uk/user_objects/}} provided by the UK \textit{Swift} Science Data Center \citep{Evans_2009}. The long term 0.3$-$10 keV XRT light curve shown in Figure \ref{fig:LCs} was also generated with the same online tool. The combined XRT spectrum was then grouped using \texttt{grppha} to ensure at least 25 counts per spectral bin, so that $\chi^2$-statistic could be applied. For spectral fitting we considered the XRT spectrum in the energy rangy 0.3$-$8 keV.

\subsubsection{IXPE}

\begin{table}
\centering
\caption{Best-fit parameters for the X-ray spectro-polarimetric analysis}
\label{tab:X-ray table}
\begingroup
\begin{tabular}{c | c | c}
\hline
\textbf{Model Component} & \textbf{Parameter} & \textbf{Value} \\
\hline
\hline
tbabs & N$_H$$^{a}$ & 2.99 $\pm$ 0.07 \\
polconst & $\Pi_X$$^{b}$ & $<$ 7.5\% \\
 & $\psi_X$ & $-$ \\
powerlaw & $\Gamma$ & 1.84 $\pm$ 0.01 \\
 & Norm$^{c}$ & 6.3 $\pm$ 0.1 \\
\hline
$\chi^2$/dof & & 1998/1990 \\
F$_{2-10 \text{keV}}^{d}$ & & -10.7 $\pm$ 0.1 \\
\hline
\multicolumn{3}{l}{\begin{minipage}{\columnwidth} 
\tablenotetext{a}{Hydrogen column density in units of $10^{21}\,\rm cm^{-2}$.}
\tablenotetext{b}{Upper limit at 99\% confidence interval.}
\tablenotetext{c}{Power-law normalization in units of $10^{-3}$ photons keV$^{-1}$ cm$^{-2}$ s$^{-1}$.}
\tablenotetext{d}{Log of flux in the energy range 2$-$10 keV in units of erg cm$^{-2}$ s$^{-1}$.}
\end{minipage}}
\end{tabular}
\endgroup
\end{table}

During the submm flaring activity, the Imaging X-Ray Polarimetry Explorer \citep[IXPE;][]{IXPE_2022} observed BL Lac for $\sim$ 10 days (2023 November 07-17; Obs. Id. 02009701) in its operating energy range, $2-8$ keV. To analyze the simultaneous IXPE data we obtained the Level 2 event files from the IXPE data archive maintained by HEASARC\footnote{\url{https://heasarc.gsfc.nasa.gov/docs/ixpe/archive/}}. The Level 2 data are already filtered, calibrated and ready for scientific use. We used \texttt{XSELECT} which is part of the \texttt{FTOOLS} software package to extract the Stokes \textit{I}, \textit{Q}, and \textit{U} source and background spectra for all 3 Detector Units (DUs). Source spectra were extracted using a circular region $60^{\prime \prime}$. Background spectra were extracted from an annular region between $150^{\prime \prime}$ and $200^{\prime \prime}$. Response files for the DUs were obtained from the calibration database (CALDB; version: 20240726). We used the IXPE-specific tool \texttt{ixpecalcarf} to generate the ancillary response files (ARFs) and modulation response files (MRFs) for all DUs. For spectro-polarimetric analysis \textit{IXPE} spectra in the energy range 2$-$8 keV were used.

\subsubsection{X-ray spectro-polarimetric analysis}

We performed a joint spectro-polarimetric fit of all the X-ray spectra in the energy range 0.3$-$78 keV using \texttt{XSPEC v12.13.1} \citep{Arnaud_Xspec}. We used a simple absorbed power-law with constant polarization model, $const \times tbabs \times (polconst \times powerlaw$), to fit the data. This model proved to be a good fit to the data with $\chi^2$/dof = 1998/1990. The photon index of the powerlaw was found to be $\Gamma$ = 1.84 $\pm$ 0.01. We left the absorbing galactic hydrogen column density, $N_H$, as a free parameter. From our fit we obtained a value of $N_H$ = 2.99 $\times 10^{21}$ cm$^{-2}$. This value is in good agreement with that obtained by \citet{Weaver_2020}. From the spectral fit we also derived that the 2$-$10 keV X-ray flux $\sim$2 $\times$ 10$^{-11}$ erg cm$^{-2}$ s$^{-1}$. This value is twice the average 2$-$10 keV flux of this source \citep[e.g.,][]{Wehrle_2016, Sahakyan_2022}. From the joint analysis of IXPE, \textit{Swift}-XRT and NuSTAR data we, however, did not find a significant detection of polarization of X-rays in the energy range 2$-$8 keV. We found only an upper limit to the polarization fraction, $\Pi_X$ $<$ 7.5\% at 99\% confidence level. Measurement of the polarization angle, $\psi_X$ was also unconstrained. 
The X-ray spectra with best-fit model and the confidence contours of parameters $\psi_X$ vs $\Pi_X$ obtained from our spectro-polarimetric fit are shown in Figure \ref{fig: X-ray pol} and best-fit parameters obtained from the X-ray spectro-polarimetric analysis are listed in Table \ref{tab:X-ray table}.

\subsection{Optical and UV Data}

\begin{table}
\centering
\caption{Perkins Telescope Observatory intra-night observation of BL Lac on Nov 13, 2023}
\label{tab:opticaldata_perkins}
\begin{tabular}{c | c | c}
\hline
\textbf{Observation Time (JD)} & \textbf{Band} & \textbf{Magnitude} \\
\hline
\hline
2460261.6494 & B & 15.115 $\pm$ 0.017 \\
2460261.6885 & B & 15.074 $\pm$ 0.015 \\
2460261.7266 & B & 15.024 $\pm$ 0.014 \\
2460261.7637 & B & 14.978 $\pm$ 0.016 \\
2460261.8047 & B & 14.900 $\pm$ 0.019 \\
\hline
2460261.6367 & V & 14.165 $\pm$ 0.015 \\
2460261.6768 & V & 14.129 $\pm$ 0.015 \\
2460261.7148 & V & 14.094 $\pm$ 0.017 \\
2460261.7529 & V & 14.040 $\pm$ 0.018 \\
2460261.7939 & V & 13.967 $\pm$ 0.017 \\
\hline
2460261.6611 & I & 12.690 $\pm$ 0.021 \\
2460261.6982 & I & 12.659 $\pm$ 0.019 \\
2460261.7354 & I & 12.616 $\pm$ 0.017 \\
2460261.7734 & I & 12.564 $\pm$ 0.016 \\
2460261.8154 & I & 12.504 $\pm$ 0.015 \\
\hline
2460261.6260 & R & 13.509 $\pm$ 0.007 \\
2460261.6279 & R & 13.507 $\pm$ 0.006 \\
2460261.6309 & R & 13.510 $\pm$ 0.005 \\
2460261.6670 & R & 13.484 $\pm$ 0.005 \\
2460261.6689 & R & 13.481 $\pm$ 0.003 \\
2460261.6719 & R & 13.480 $\pm$ 0.004 \\
2460261.7041 & R & 13.449 $\pm$ 0.005 \\
2460261.7070 & R & 13.446 $\pm$ 0.005 \\
2460261.7090 & R & 13.446 $\pm$ 0.005 \\
2460261.7422 & R & 13.402 $\pm$ 0.004 \\
2460261.7441 & R & 13.399 $\pm$ 0.005 \\
2460261.7471 & R & 13.397 $\pm$ 0.006 \\
2460261.7832 & R & 13.332 $\pm$ 0.004 \\
2460261.7861 & R & 13.331 $\pm$ 0.004 \\
2460261.7881 & R & 13.326 $\pm$ 0.005 \\
\hline
\end{tabular}
\end{table}

\subsubsection{\textit{Swift} UVOT}

BL Lac was observed with \textit{Swift} Ultraviolet and Optical Telescope \citep[UVOT;][]{Roming_2005_UVOT} alongside the XRT observations. We use the data from the UVOT observations performed on 12$-$14 November, 2023 since this time period is simultaneous with the \textit{Perkins} Telescope observations (described in the next subsection) and during the peak of the submm flare. UVOT observed the source in six filters (\textit{UVW2}, \textit{UVM2}, \textit{UVW1}, $U$, $B$ and $V$). We processed the data using HEASoft v6.26. We subsequently performed aperture photometry on the processed images using the tool \texttt{UVOTSOURCE} to obtain the fluxes in each filter. For aperture photometry a source region of 5$^{\prime \prime}$ was used and a nearby source-free background region of 20$^{\prime \prime}$ was used. The flux values obtained were corrected for extinction using $E(B-V) = 0.29$ as given in the NASA/IPAC Extragalactic Database\footnote{\url{https://ned.ipac.caltech.edu/}} by \cite{Schlafly_Finkbeiner_2011}. Extinction correction was performed using the extinction relation of \citet{Cardelli_1989}. After extinction correction, we subtracted the appropriate contribution of the host galaxy from the extinction corrected fluxes following the values given by \citet{Raiteri_2013}. 

\begin{table}
\centering
\caption{\textbf{\textit{Swift} UVOT and Perkins data}}
\label{tab:uv_opticaldata}
\begingroup
\begin{tabular}{c | c | c | c}
\hline
\textbf{Facilities} & \textbf{Time$^{a}$} & \textbf{Band} & \textbf{$\nu F_{\nu}$$^{b}$} \\
\hline
\hline
\multirow{3}{*}{UVOT} & $60260 - 60262$ & UVW2 & $4.28\pm0.14$ \\
 & $60260 - 60262$ & UVM2 & $6.64\pm0.25$ \\
 & $60260 - 60262$ & UVW1 & $5.27\pm0.11$ \\
\hline
\multirow{3}{*}{UVOT} & $60260 - 60262$ & U & $6.18\pm0.10$ \\
 & $60260 - 60262$ & B & $6.95\pm0.13$ \\
 & $60260 - 60262$ & V & $10.73\pm0.19$ \\
\hline
\multirow{4}{*}{Perkins} & 60261 & B & $8.86\pm0.89$ \\
 & 60261 & V & $10.14\pm0.42$ \\
 & 60261 & R & $9.21\pm0.22$ \\
 & 60261 & I & $9.89\pm1.07$ \\
\hline
\multicolumn{4}{l}{\begin{minipage}{\columnwidth}
\tablenotetext{a}{observation time in MJD}
\tablenotetext{b}{$\nu F_{\nu}$ in units of $10^{-11} \times \text{erg cm}^{-2}\text{s}^{-1}$}
\end{minipage}}
\end{tabular}
\endgroup
\end{table}

\subsubsection{Optical Data: Perkins Telescope Observatory}

On MJD 60261 (2023 Nov 13) BL Lacertae was observed with the 1.83-meter telescope located at the Perkins Telescope Observatory (PTO) in Arizona, USA, owned and operated by Boston University. A total of 15 observations of the source in the $R$ band and 5 observations in each of the $B$, $V$, and $I$ bands were conducted throughout the night, which are listed in Table \ref{tab:opticaldata_perkins}. We have used the average of the intra-night data  at all four optical bands in our SED.

We use both UVOT and Perkins observation for each individual bands separately in the SED which are listed in Table \ref{tab:uv_opticaldata}.

\subsection{Submillimeter Array (SMA)}
The Submillimeter Array (SMA) is an 8-element radio interferometer situated atop Maunakea in Hawaii \citep{Ho_2004}. During the period spanning 2023 mid-October to November, SMA observed the flaring event of BL Lacertae. The intensity of the flaring event persisted for weeks, and on November 14 (MJD 60262), the flux density reached 21 Jy, exceeding, by over 30\%, the flux levels of any previous flare by BL Lac in the 20-year light curve measured by the SMA \citep{2023ATel16340....1G}. Over this duration, the source underwent regular monitoring by SMA, and a total of 7 observations were recorded. We have taken the average of all those seven observations by SMA in the $1.4 - 1.1$ mm band. All available data from the SMA Observer Center\footnote{\url{http://sma1.sma.Hawaii.edu/callist/callist.html}} for the month of November have been utilized in our SED analysis. The long-term SMA light curve in the $1.4 - 1.1$ mm band shown in Figure \ref{fig:LCs} is also obtained from the SMA Observer Center.

\begin{table}
\centering
\caption{VLBA 15 GHz \& 43 GHz observation during the flaring episode in November, 2023. VLBA I is the total cleaned Stokes $I$ flux density, VLBA $P$ is the total cleaned linearly polarized flux density, total fractional linear polarization ($P$), and integrated electric vector position angle (EVPA) are shown in the table.}
\label{tab:VLBAdata}
\begin{tabular}{c | c | c | c | c}
\hline
\textbf{Epoch} & \textbf{VLBA I} & \textbf{VLBA P} & \textbf{P} & \textbf{EVPA} \\
 & \textbf{(mJy)} & \textbf{(mJy)} & \textbf{(\%)} & \textbf{(deg.)} \\
\hline
\hline
14 Nov (15GHz) & 10964 & 231 & 2.10 & 106 \\
\hline
06 Nov (43GHz) & 15145 & 518.9 & 3.42 & -24 \\
25 Nov (43GHz) & 16278 & 369 & 2.26 & 53 \\
\hline
\end{tabular}
\end{table}

\subsection{VLBA Radio Observation}

In blazars, mm/submm flares typically precede outbursts in the radio band. In this case also, on 2023 November, a large radio flux was observed in the Very Long Baseline Array (VLBA) 15 GHz MOJAVE survey\footnote{\url{https://www.cv.nrao.edu/MOJAVE/sourcepages/2200+420.shtml}} \citep{Lister_mojave} and 43 GHz monitoring by the BEAM-ME survey\footnote{\url{https://www.bu.edu/blazars/VLBA_GLAST/bllac.html}} shown in Table \ref{tab:VLBAdata}. We utilized those data in our SED analysis. Long term VLBA light curve from 15 GHz MOJAVE survey archive is shown in Figure \ref{fig:LCs}.

\section{SED Modeling} \label{sec:SED_modeling}

We model the broadband SED of BL Lac in the leptonic scenario. We use the publicly available code \textit{JetSeT} (Version 1.3.0)\footnote{\url{https://jetset.readthedocs.io/en/latest/index.html}} \citep{Tramacere_2009,Tramacere_2011,JetSeT_2020}, which models synchrotron emission as well as SSC and EC processes. We fix certain parameters of the model to values constrained by previous observations or based on functional dependence established in the literature (listed in Table \ref{tab:SED_Results}). For example, the disk luminosity is fixed at $2.19 \times 10^{43} \, \text{erg} \, \text{s}^{-1}$ \citep[VizieR Online Data Catalog: Optical spectroscopy of \textit{Fermi} blazars;][]{Paliya_VizieR_2021}. The distance of the BLR from the central engine depends on the luminosity of the accretion disk according to the expressions given by \cite{Kaspi_2007}:
\begin{equation}
    R_{\text{BLR,in}} = 3 \times 10^{17} \left(\frac{L_{\text{disk}}}{10^{46}}\right)^{0.5} cm
\end{equation}

\begin{equation}
    R_{\text{BLR,out}} = R_{\text{BLR,in}} \times 1.1
\end{equation}

The distance of IR-torus from the central engine depends on the luminosity of the accretion disk according to the expression given by \cite{Cleary_2007}:
\begin{equation}
        R_{\text{DT}} = 2 \times 10^{19} \left(\frac{L_{\text{disk}}}{10^{46}}\right)^{0.5} cm
\end{equation}

Here we consider the emission region fills the entire cross section of the relativistic jet. Thus the jet opening angle ($\theta_{\text{open}}$), distance ($R_{\text{H}}$) and size of the emission region ($R$) are linked by the expressions \[ R = \tan(\theta_{\text{open}}) \cdot R_H \] We fix $\theta_{\text{open}}$ at $3^\circ$ based on existing literature, which is also consistent with the average jet opening angles observed in BL Lac objects \citep{Pushkarev_2009,Pushkarev_2017}. $R$ and $R_{\text{H}}$ are initially chosen based on those reported in the literature \citep{Sahakyan_2022, Zahir_2023} and are kept free during the model fitting process within a certain limit of acceptable values. The accretion disk is modeled as a multi-color blackbody, and the BLR and torus as monochromatic emitters. We assume the observed broadband emission is radiated by a population of relativistic electrons following an energy distribution defined by a power-law with an exponential cutoff given by:
\begin{equation}
N(\gamma) = N_e \gamma^{-p} \exp\left(-\frac{\gamma}{\gamma_{\text{cut}}}\right),
\end{equation}
where, $\gamma_{\text{min}} \leq \gamma \leq \gamma_{\text{max}}$ and $\gamma_{\text{min}}, \gamma_{\text{max}}$ and $\gamma_{\text{cut}}$ are the minimum, maximum and cut-off energies of the electron energy distribution and, $p$ is the spectral index of the power-law distribution.

\begin{figure}[htb!]
\centering
\includegraphics[width=\columnwidth]{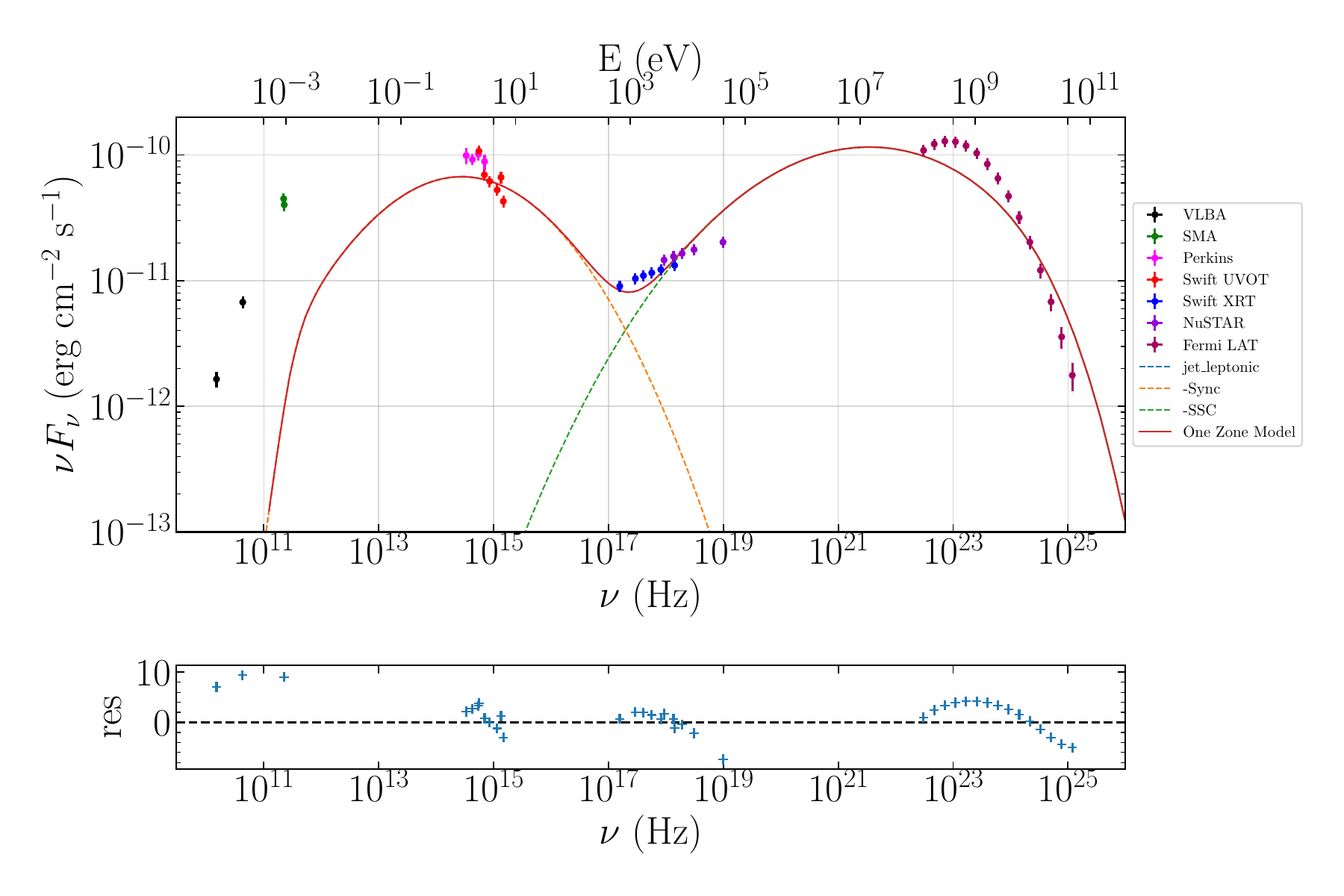}
\caption{Broadband SED fitting of BL Lacertae during November 2023 with a one-zone leptonic model using \textit{JetSeT}. It indicates that the usual one-zone leptonic model with SSC is inadequate to fit the data with the high flux around submm wavelength.}
\label{fig: one_zone SED}
\end{figure}

\begin{figure*}[htb!]
\centering
\includegraphics[width=\textwidth]{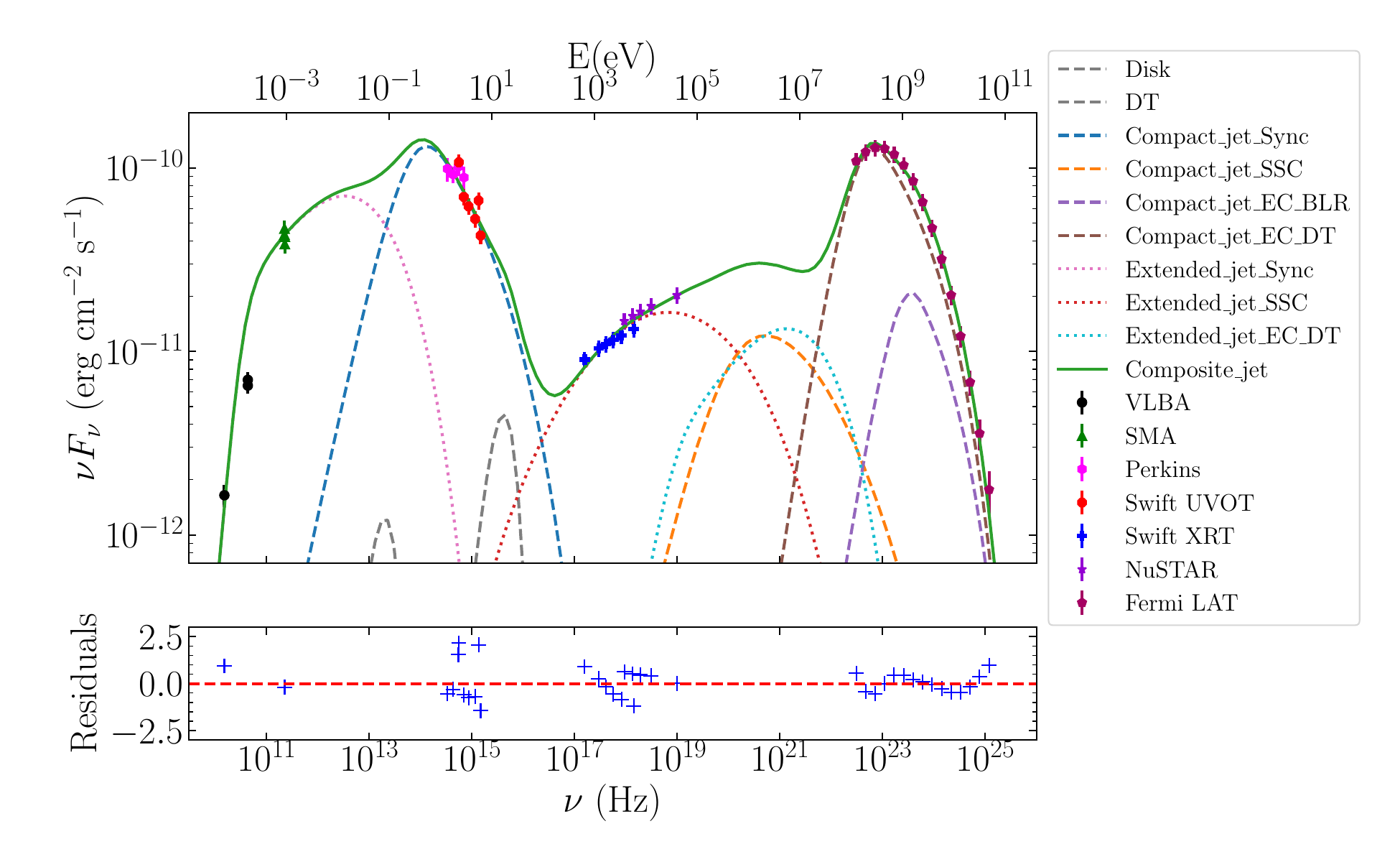}
\caption{Broadband SED fitting of BL Lacertae during November 2023. The SED fit is performed using \textit{JetSeT} with a two-zone leptonic model scenario. The dashed-lines are the components from the near emission zone, the dotted-lines indicate the components from the far emission zone, and the solid line indicates the sum of all components.}
\label{fig:SED}
\end{figure*}

\begin{table*}
\centering
\caption{Best-fit parameters of near zone and far zone. The near emission zone is outside the BLR \& the far emission zone is outside the torus. The parameters at the bottom (without the uncertainties) are frozen during the model fitting.}
\label{tab:SED_Results}
\begin{tabular}{l | c c | c c}
\hline
\multirow{2}{*}{\textbf{Parameters}} & \multicolumn{2}{c |}{\textbf{Near Zone}} & \multicolumn{2}{c}{\textbf{Far zone}} \\
& \textbf{bestfit value} & \textbf{error} & \textbf{bestfit value} & \textbf{error} \\
\hline
\hline
Emission region size (R) cm & $2.27 \times 10^{15}$ & -- & $1.23 \times 10^{17}$ & -- \\
Emission region distance ($R_{\text{H}}$) cm & $3.70 \times 10^{16}$ & $5.96 \times 10^{11}$ & $1.99 \times 10^{18}$ & $5.60 \times 10^{16}$ \\
Magnetic field (B) Gauss & $0.894$ & $0.003$ & $0.0272$ & $0.0002$ \\
Bulk Lorentz Factor ($\Gamma$) & 28.8 & 0.3 & 8.87 & 0.03 \\
Minimum Lorentz factor ($\gamma_{\text{min}}$) & 698.0 & 9.0 & 104.0 & 1.0 \\
Maximum Lorentz factor ($\gamma_{\text{max}}$) & $1.09 \times 10^{6}$ & $3.59 \times 10^{3}$ & $8.12 \times 10^{5}$ & $2.08 \times 10^{3}$ \\
Turn over energy ($\gamma_{\text{cut}}$) & $4.35 \times 10^{3}$ & $2.13 \times 10^{2}$ & $1.73 \times 10^{3}$ & $5.19 \times 10^{1}$ \\
Low-energy spectral slope (p) & 3.452 & 0.004 & 2.0304 & 0.0001 \\
\hline
Inner radius of BLR ($R_{\text{BLR,in}}$) cm & \multicolumn{2}{c |}{$1.40 \times 10^{16}$} & \multicolumn{2}{c}{$1.40 \times 10^{16}$} \\
Outer radius of BLR ($R_{\text{BLR,out}}$) cm & \multicolumn{2}{c |}{$1.54 \times 10^{16}$} & \multicolumn{2}{c}{$1.54 \times 10^{16}$} \\
Torus temperature ($T_{\text{DT}}$) K & \multicolumn{2}{c |}{$3.32 \times 10^{2}$} & \multicolumn{2}{c}{$3.32 \times 10^{2}$} \\
Size of the torus ($R_{\text{DT}}$) cm & \multicolumn{2}{c |}{$9.35 \times 10^{17}$} & \multicolumn{2}{c}{$9.35 \times 10^{17}$} \\
$\tau_{\text{DT}}$ & \multicolumn{2}{c |}{$9.20 \times 10^{-1}$} & \multicolumn{2}{c}{$9.20 \times 10^{-1}$} \\
Accretion disk luminosity ($L_{\text{disk}}$) $\text{erg s}^{-1}$ (fixed) & \multicolumn{2}{c |}{$2.19 \times 10^{43}$} & \multicolumn{2}{c}{$2.19 \times 10^{43}$} \\
Accretion disk temperature ($T_{\text{Disk}}$) K & \multicolumn{2}{c |}{$1.16 \times 10^{5}$} & \multicolumn{2}{c}{$1.16 \times 10^{5}$} \\
\hline
\end{tabular}
\end{table*}

\subsection{Results of the SED analysis}
While modeling the SED, we first consider a standard leptonic model, in which the lower energy emission, from radio to optical, is modeled by synchrotron radiation from the above electron distribution and the X-ray--$\gamma$-ray emission is fit by the SSC and/or EC process due to the same electrons. We apply a systematic error of $10\%$ evenly across the entire multi-wavelength dataset \citep{Zahir_2023, Rajguru_2024}. 
\textit{JetSeT} uses Minuit and Markov-Chain Monte-Carlo (MCMC) optimization methods to obtain the best-fit parameters. However, the best-fit model found with this scenario is not satisfactory. In particular, it cannot reproduce the high flux around submm wavelengths (see Figure \ref{fig: one_zone SED}).

Therefore, we consider a leptonic model with two emission regions containing separate electron distributions, located at two positions along the jet nearer to and farther away from the central engine. The near and far emission regions are assumed to be spherical blobs of radius $R_{1}$ and $R_{2}$, respectively, where the magnetic fields are $B_{1}$ and $B_{2}$ with $B_{1} > B_{2}$. For the lower energy part of the SED, we consider two different synchrotron components due to the two separate distributions of electrons. The synchrotron emission from the near zone fits the optical part of the SED. Emission through the SSC process from the near zone contributes to the hard X-ray to $\gamma$-ray frequencies while the GeV energy region is fit by the EC components with seed photons from BLR and torus. On the other hand, the submm emission is modeled by the far-zone synchrotron component and the far-zone SSC emission is used to model the X-ray portion of the SED (see Figure \ref{fig:SED}). The used and derived best-fit model parameters of near and far zone for the above model are listed in Table \ref{tab:SED_Results}.

The 0.3–100 keV X-ray emission is mostly generated by the SSC process at the emitting zone farther downstream along the jet. There are also minor contributions of the SSC process from the compact jet and external Compton (EC) scattering from the extended jet. EC scattering of the torus seed photons, along with SSC in the compact jet, dominates the emission in the MeV range. EC scattering from the BLR and the torus in the compact jet leads to GeV emission. In the extended jet, the contribution from the EC scattering of the BLR seed photons is negligible because the extended zone is further downstream along the jet and hence more exposed to the seed photon field of the torus than that of the BLR.

The above model with two separate distributions of electrons located at two emission regions nearer and farther from the central engine provide a satisfactory fit to the broadband SED constructed using data at radio--submm-optical--X-ray--GeV energies. While we have constructed and modeled the SED the light curves are not sampled well enough to suitably constrain the cross-correlation time delays between different emission components. However, one can find in Figure \ref{fig:LCs} that there may be a hint of correlation between submm and X-ray light curves. On the other hand, the GeV variability does not exhibit a strong correlation with that at submm or X-ray wavelengths. Both of those effects are expected because the GeV emission, as per our interpretation, is from the compact jet while the submm and X-rays are from the distant emission region. We note that, location of GeV emission region to be a few pc upstream of the mm-radio emission region has been inferred by other authors in different sources, e.g., \citet{2015IAUS..313...39L} for the blazars 3C 273 and 3C 279. Our findings are consistent with those results.

However, the above model has a larger number of free parameters and hence a better fit is expected although it does not necessarily indicate a better description of the ongoing emission process. But we can use simultaneous X-ray polarization information to probe the above physical scenario. IXPE had no detection of X-ray polarization and only provided an upper limit of $< 7.5\%$ for the polarization fraction. Previously, significantly larger value of X-ray polarization has been detected by IXPE for several HSP blazars \citep[see, \textit{e.g.},][]{Di_Gesu_2022, Marscher_book_2023, Kouch_2024, Liodakis_2022Nature, Di_Gesu_2023NatAs, Ehlert_2023}. Moderately high value of polarization fraction is also observed at the optical and radio wave bands during large outbursts in many blazars \citep{Sasada_2014, Fraija_2017, Ueura_2017, Jorstad_2022_nature, Bachev_2023}. The absence of significant polarization at that epoch of very large submm flare along with high values of optical emission may support the physical scenario indicated by our best-fit model that the X-ray part of the SED is dominated by the far-zone SSC contribution, and the polarization is reduced by the Comptonization process \citep{Krawczynski_2012, Peirson_2019}. We note that in other HSP blazars, in which high value of polarization fraction had been observed by IXPE, the X-ray emission was dominated by the synchrotron process, which can lead to strongly polarized emission. 
We note that the polarization fractions at 15 GHz and 43 GHz radio wavelengths, as shown in Table \ref{tab:VLBAdata}, are not too high. That may imply that even the synchrotron emission from the far zone or extended jet is not highly polarized. However, high value of polarization may still be found at the optical wave band, which is from the near zone or compact jet.  

\section{Discussion} \label{sec:Discussion}
BL Lac was observed by IXPE during three epochs in 2022. In 2022 May and July, during a low X-ray state, \cite{Middei_2023} could only determine an upper limit on the X-ray polarization fraction ($\Pi_X$) and the polarization angle was unconstrained although the optical polarization fraction ($\Pi_O$) was as high as 17\% while that in the radio/mm band ($\Pi_R$) was 9\%. In order to explain it, the authors suggested that the X-rays are generated in the SSC process, in which a decrement in the polarization fraction by a factor $\sim 3$ compared to the synchrotron seed photons is theoretically expected \citep{Krawczynski_2012, Peirson_2019}. In addition, \cite{Liodakis_2019} conclude that if external photons contribute seed photons to the IC process, polarization fraction will also decrease. Later, in 2022 November, during an outburst of BL Lacertae at X-ray and $\gamma$-ray energies, $\Pi_{2-4 \text{keV}}$ = $21.7^{+5.6}_{-7.9}$\% was determined with 99\% confidence level while the polarization fraction was $5\%-8\%$ in radio/mm bands and $\Pi_O \sim 10\%$ \citep{Perison_2023}. The authors suggested that the high level of polarization at softer X-rays indicate that the high-energy tail of the synchrotron component extends to soft X-rays during the outburst. Synchrotron radiation at X-ray energies is produced by the highest energy electrons, which may be confined in a smaller region with highly ordered magnetic field giving rise to the large polarization fraction \citep[e.g.,][]{Marscher_2024}. The optical emission, also due to synchrotron process, is generated by relatively lower energy electrons located in a larger region leading to a lower polarization value. We note that if the observed emission is produced in $N$ turbulent cells or regions having different orientation of magnetic field then the observed polarization fraction will decrease by a factor $N^{1/2}$ compared to the polarization fraction of the individual cell's emission \citep{Marscher_2014, Jorstad_2007}. In all of the above cases, the lack of high value of polarization at harder X-rays favors the leptonic scenario because if proton synchrotron radiation contribute significantly to the X-ray emission then a higher polarization fraction, than observed, is expected. 

Broadband SEDs of blazars are often fit satisfactorily by the nonthermal emission from a single distribution of electrons. However, in certain epochs, blazars have been shown to exhibit multi-wavelength emission that cannot be explained by the same without having very unrealistic physical parameters. In those cases, a second distribution of electrons are invoked in order to fit the broadband SED. In the epoch studied here, BL Lac underwent a very high-amplitude outburst at submm band, which, along with its simultaneous observations at optical--X-ray--GeV energies, cannot be explained by the radiation from a single leptonic distribution. Previously, it has been observed by, e.g., \citet{Sahakyan_2022}, that more than one distribution of emitting particles are required to fit the SED during bright flares in VHE $\gamma$-rays or X-rays. During the epoch discussed in this work, BL Lac is brighter at the 2$-$10 keV energy band than its average by a factor $\sim 2$ and the SED can be fit satisfactorily with two electron distributions.

In this epoch, we find only an upper limit on the X-ray polarization while the radio polarization has moderate value. From the X-ray spectrum, it is clear that the X-rays, even at the softer energies, are dominated by the SSC process, which may lead to low polarization. However, in addition to the SSC origin, the low X-ray polarization may also be due to its production in a region with a higher level of disorder in the magnetic field. Low polarization fraction is detected in the radio emission, which is also generated in the far zone or extended jet. That implies the polarization level at mm-wave band will also not be too high. Otherwise, polarization of the SSC X-rays produced from the mm-wave synchrotron seed photons could have been large enough for detection. On the other hand, optical synchrotron emission, produced by the highest energy electrons in the near zone possibly containing a more ordered magnetic field, may exhibit a high degree of polarization. This result demonstrates that multi-wavelength polarization observations, including that at X-ray energies, can provide crucial constraints in favoring certain emission models and physical parameters, which may help breaking the degeneracy that often affects the conclusions drawn from blazar SED modeling. 


\section{Acknowledgments}
We thank the anonymous referee for useful comments that helped in improving the manuscript. RC and AS thank ISRO for support under the \textit{AstroSat} archival data utilization program, and IUCAA for their hospitality and usage of their facilities during their stay at different times as part of the university associateship program. MK acknowledges financial support from SERB through the POWER Fellowship (SPF/2022/000084), and from Washington University in St.\ Louis through the Arts \& Sciences Fellowship. RC thanks Presidency University for support under the Faculty Research and Professional Development (FRPDF) Grant, and acknowledges financial support from SERB through a SURE grant (SUR/2022/001503). AM thanks Andrea Tramacere for assistance with the \textit{JetSeT} code. This research has made use of the NASA/IPAC Extragalactic Database (NED), which is funded by the National Aeronautics and Space Administration and operated by the California Institute of Technology; data from the MOJAVE database that is maintained by the MOJAVE team; VLBA data from the VLBA-BU Blazar Monitoring Program (BEAM-ME and VLBA-BU-BLAZAR) funded by NASA through a Fermi GI program; observations conducted using the 1.83-m Perkins Telescope Observatory (PTO) in Arizona (USA), which is owned and operated by Boston University. We thank Svetlana Jorstad for sharing the optical photometric data with us. Submillimeter Array (SMA) data used in this work have been provided to us on request. The SMA is a joint project between the Smithsonian Astrophysical Observatory and the Academia Sinica Institute of Astronomy and Astrophysics and is funded by the Smithsonian Institution and the Academia Sinica located near the summit of Maunakea in Hawaii. We also recognize that Maunakea is a culturally important site for the indigenous Hawaiian people; we are privileged to study the cosmos from its summit. 

\section{Data availability} \label{sec:data_avl}

The data used in this study are publicly available in \textit{Fermi}, \textit{NuSTAR}, \textit{Swift}, IXPE, SMA \& VLBA archives.

\vspace{5mm}
\facilities{\textit{Fermi}-LAT, NuSTAR, \textit{Swift} (XRT and UVOT), IXPE, Perkins, SMA, VLBA}

\software{ JetSeT \citep{Tramacere_2009, Tramacere_2011, JetSeT_2020}, Astropy \citep{2013A&A...558A..33A,2018AJ....156..123A}, Xspec \citep{Arnaud_Xspec,Dorman_Xspec,Dorman_2003}, Fermitools \citep{Fermitools}
          }


\bibliography{main}

\begin{thebibliography}{}
\expandafter\ifx\csname natexlab\endcsname\relax\def\natexlab#1{#1}\fi
\providecommand{\url}[1]{\href{#1}{#1}}
\providecommand{\dodoi}[1]{doi:~\href{http://doi.org/#1}{\nolinkurl{#1}}}
\providecommand{\doeprint}[1]{\href{http://ascl.net/#1}{\nolinkurl{http://ascl.net/#1}}}
\providecommand{\doarXiv}[1]{\href{https://arxiv.org/abs/#1}{\nolinkurl{https://arxiv.org/abs/#1}}}

\bibitem[{Abdo {et~al.}(2010)Abdo, Ackermann, Agudo, Ajello, Aller, Aller,
  Angelakis, Arkharov, Axelsson, Bach, Baldini, Ballet, Barbiellini, Bastieri,
  Baughman, Bechtol, Bellazzini, Benitez, Berdyugin, Berenji, Blandford, Bloom,
  Boettcher, Bonamente, Borgland, Bregeon, Brez, Brigida, Bruel, Burnett,
  Burrows, Buson, Caliandro, Calzoletti, Cameron, Capalbi, Caraveo, Carosati,
  Casandjian, Cavazzuti, Cecchi, Çelik, Charles, Chaty, Chekhtman, Chen,
  Chiang, Chincarini, Ciprini, Claus, Cohen-Tanugi, Colafrancesco, Cominsky,
  Conrad, Costamante, Cutini, D'ammando, Deitrick, D'Elia, Dermer, de~Angelis,
  de~Palma, Digel, Donnarumma, do~Couto~e Silva, Drell, Dubois, Dultzin,
  Dumora, Falcone, Farnier, Favuzzi, Fegan, Focke, Forné, Fortin, Frailis,
  Fuhrmann, Fukazawa, Funk, Fusco, Gómez, Gargano, Gasparrini, Gehrels,
  Germani, Giebels, Giglietto, Giommi, Giordano, Giuliani, Glanzman, Godfrey,
  Grenier, Gronwall, Grove, Guillemot, Guiriec, Gurwell, Hadasch, Hanabata,
  Harding, Hayashida, Hays, Healey, Heidt, Hiriart, Horan, Hoversten, Hughes,
  Itoh, Jackson, Jóhannesson, Johnson, Johnson, Jorstad, Kadler, Kamae,
  Katagiri, Kataoka, Kawai, Kennea, Kerr, Kimeridze, Knödlseder, Kocian,
  Kopatskaya, Koptelova, Konstantinova, Kovalev, Kovalev, Kurtanidze, Kuss,
  Lande, Larionov, Latronico, Leto, Lindfors, Longo, Loparco, Lott, Lovellette,
  Lubrano, Madejski, Makeev, Marchegiani, Marscher, Marshall, Max-Moerbeck,
  Mazziotta, McConville, McEnery, Meurer, Michelson, Mitthumsiri, Mizuno,
  Moiseev, Monte, Monzani, Morselli, Moskalenko, Murgia, Nestoras, Nilsson,
  Nizhelsky, Nolan, Norris, Nuss, Ohsugi, Ojha, Omodei, Orlando, Ormes,
  Osborne, Ozaki, Pacciani, Padovani, Pagani, Page, Paneque, Panetta, Parent,
  Pasanen, Pavlidou, Pelassa, Pepe, Perri, Pesce-Rollins, Piranomonte, Piron,
  Pittori, Porter, Puccetti, Rahoui, Rainò, Raiteri, Rando, Razzano, Reimer,
  Reimer, Reposeur, Richards, Ritz, Rochester, Rodriguez, Romani, Ros, Roth,
  Roustazadeh, Ryde, Sadrozinski, Sadun, Sanchez, Sander, Parkinson, Scargle,
  Sellerholm, Sgrò, Shaw, Sigua, Siskind, Smith, Smith, Spandre, Spinelli,
  Starck, Stevenson, Stratta, Strickman, Suson, Tajima, Takahashi, Takahashi,
  Takalo, Tanaka, Thayer, Thayer, Thompson, Tibaldo, Torres, Tosti, Tramacere,
  Uchiyama, Usher, Vasileiou, Verrecchia, Vilchez, Villata, Vitale, Waite,
  Wang, Winer, Wood, Ylinen, Zensus, Zhekanis, \& Ziegler}]{Abdo_2010}
Abdo, A.~A., Ackermann, M., Agudo, I., {et~al.} 2010, The Astrophysical
  Journal, 716, 30, \dodoi{10.1088/0004-637X/716/1/30}

\bibitem[{{Abdo} {et~al.}(2011{\natexlab{a}}){Abdo}, {Ackermann}, {Ajello},
  {Baldini}, {Ballet}, {Barbiellini}, {Bastieri}, {Bechtol}, {Bellazzini},
  {Berenji}, {Blandford}, {Bloom}, {Bonamente}, {Borgland}, {Bouvier},
  {Bregeon}, {Brez}, {Brigida}, {Bruel}, {Buehler}, {Buson}, {Caliandro},
  {Cameron}, {Cannon}, {Caraveo}, {Carrigan}, {Casandjian}, {Cavazzuti},
  {Cecchi}, {{\c{C}}elik}, {Charles}, {Chekhtman}, {Chiang}, {Ciprini},
  {Claus}, {Cohen-Tanugi}, {Conrad}, {Cutini}, {de Angelis}, {de Palma},
  {Dermer}, {Silva}, {Drell}, {Dubois}, {Dumora}, {Escande}, {Favuzzi},
  {Fegan}, {Finke}, {Focke}, {Fortin}, {Frailis}, {Fuhrmann}, {Fukazawa},
  {Fukuyama}, {Funk}, {Fusco}, {Gargano}, {Gasparrini}, {Gehrels},
  {Georganopoulos}, {Germani}, {Giebels}, {Giglietto}, {Giommi}, {Giordano},
  {Giroletti}, {Glanzman}, {Godfrey}, {Grenier}, {Guiriec}, {Hadasch},
  {Hayashida}, {Hays}, {Horan}, {Hughes}, {J{\'o}hannesson}, {Johnson},
  {Johnson}, {Kadler}, {Kamae}, {Katagiri}, {Kataoka}, {Kn{\"o}dlseder},
  {Kuss}, {Lande}, {Latronico}, {Lee}, {Longo}, {Loparco}, {Lott},
  {Lovellette}, {Lubrano}, {Madejski}, {Makeev}, {Max-Moerbeck}, {Mazziotta},
  {McEnery}, {Mehault}, {Michelson}, {Mitthumsiri}, {Mizuno}, {Monte},
  {Monzani}, {Morselli}, {Moskalenko}, {Murgia}, {Nakamori}, {Naumann-Godo},
  {Nishino}, {Nolan}, {Norris}, {Nuss}, {Ohsugi}, {Okumura}, {Omodei},
  {Orlando}, {Ormes}, {Ozaki}, {Paneque}, {Panetta}, {Parent}, {Pavlidou},
  {Pearson}, {Pelassa}, {Pepe}, {Pesce-Rollins}, {Pierbattista}, {Piron},
  {Porter}, {Rain{\`o}}, {Rando}, {Razzano}, {Readhead}, {Reimer}, {Reimer},
  {Reyes}, {Richards}, {Ritz}, {Roth}, {Sadrozinski}, {Sanchez}, {Sander},
  {Sgr{\`o}}, {Siskind}, {Smith}, {Spandre}, {Spinelli}, {Stawarz},
  {Stevenson}, {Strickman}, {Suson}, {Takahashi}, {Takahashi}, {Tanaka},
  {Thayer}, {Thayer}, {Thompson}, {Tibaldo}, {Torres}, {Tosti}, {Tramacere},
  {Troja}, {Usher}, {Vandenbroucke}, {Vasileiou}, {Vianello}, {Vilchez},
  {Vitale}, {Waite}, {Wang}, {Wehrle}, {Winer}, {Wood}, {Yang}, {Yatsu},
  {Ylinen}, {Zensus}, {Ziegler}, {Fermi LAT Collaboration}, {Aleksi{\'c}},
  {Antonelli}, {Antoranz}, {Backes}, {Barrio}, {Becerra Gonz{\'a}lez},
  {Bednarek}, {Berdyugin}, {Berger}, {Bernardini}, {Biland}, {Blanch}, {Bock},
  {Boller}, {Bonnoli}, {Bordas}, {Borla Tridon}, {Bosch-Ramon}, {Bose},
  {Braun}, {Bretz}, {Camara}, {Carmona}, {Carosi}, {Colin}, {Colombo},
  {Contreras}, {Cortina}, {Covino}, {Dazzi}, {de Angelis}, {De Cea del Pozo},
  {Delgado Mendez}, {De Lotto}, {De Maria}, {De Sabata}, {Diago Ortega},
  {Doert}, {Dom{\'\i}nguez}, {Dominis Prester}, {Dorner}, {Doro}, {Elsaesser},
  {Ferenc}, {Fonseca}, {Font}, {Garc{\'\i}a L{\'o}pez}, {Garczarczyk}, {Gaug},
  {Giavitto}, {Godinovi}, {Hadasch}, {Herrero}, {Hildebrand},
  {H{\"o}hne-M{\"o}nch}, {Hose}, {Hrupec}, {Jogler}, {Klepser},
  {Kr{\"a}henb{\"u}hl}, {Kranich}, {Krause}, {La Barbera}, {Leonardo},
  {Lindfors}, {Lombardi}, {L{\'o}pez}, {Lorenz}, {Majumdar}, {Makariev},
  {Maneva}, {Mankuzhiyil}, {Mannheim}, {Maraschi}, {Mariotti}, {Mart{\'\i}nez},
  {Mazin}, {Meucci}, {Miranda}, {Mirzoyan}, {Miyamoto}, {Mold{\'o}n},
  {Moralejo}, {Nieto}, {Nilsson}, {Orito}, {Oya}, {Paoletti}, {Paredes},
  {Partini}, {Pasanen}, {Pauss}, {Pegna}, {Perez-Torres}, {Persic}, {Peruzzo},
  {Pochon}, {Prada}, {Prada Moroni}, {Prandini}, {Puchades}, {Puljak},
  {Reichardt}, {Rhode}, {Rib{\'o}}, {Rico}, {Rissi}, {R{\"u}gamer}, {Saggion},
  {Saito}, {Saito}, {Salvati}, {S{\'a}nchez-Conde}, {Satalecka}, {Scalzotto},
  {Scapin}, {Schultz}, {Schweizer}, {Shayduk}, {Shore}, {Sierpowska-Bartosik},
  {Sillanp{\"a}{\"a}}, {Sitarek}, {Sobczynska}, {Spanier}, {Spiro}, {Stamerra},
  {Steinke}, {Storz}, {Strah}, {Struebig}, {Suric}, {Takalo}, {Tavecchio},
  {Temnikov}, {Terzi{\'c}}, {Tescaro}, {Teshima}, {Vankov}, {Wagner},
  {Weitzel}, {Zabalza}, {Zandanel}, {Zanin}, {MAGIC Collaboration}, {Villata},
  {Raiteri}, {Aller}, {Aller}, {Chen}, {Jordan}, {Koptelova}, {Kurtanidze},
  {L{\"a}hteenm{\"a}ki}, {McBreen}, {Larionov}, {Lin}, {Nikolashvili},
  {Reinthal}, {Angelakis}, {Capalbi}, {Carrami{\~n}ana}, {Carrasco}, {Cassaro},
  {Cesarini}, {Falcone}, {Gurwell}, {Hovatta}, {Kovalev}, {Kovalev},
  {Krichbaum}, {Krimm}, {Lister}, {Moody}, {Maccaferri}, {Mori}, {Nestoras},
  {Orlati}, {Pace}, {Pagani}, {Pearson}, {Perri}, {Piner}, {Ros}, {Sadun},
  {Sakamoto}, {Tammi}, \& {Zook}}]{Abdo_2011_mrk421}
{Abdo}, A.~A., {Ackermann}, M., {Ajello}, M., {et~al.} 2011{\natexlab{a}},
  \apj, 736, 131, \dodoi{10.1088/0004-637X/736/2/131}

\bibitem[{{Abdo} {et~al.}(2011{\natexlab{b}}){Abdo}, {Ackermann}, {Ajello},
  {Antolini}, {Baldini}, {Ballet}, {Barbiellini}, {Bastieri}, {Bechtol},
  {Bellazzini}, {Berenji}, {Blandford}, {Bonamente}, {Borgland}, {Bregeon},
  {Brez}, {Brigida}, {Bruel}, {Buehler}, {Buson}, {Caliandro}, {Cameron},
  {Cannon}, {Caraveo}, {Carrigan}, {Casandjian}, {Cecchi}, {{\c{C}}elik},
  {Charles}, {Chekhtman}, {Cheung}, {Chiang}, {Ciprini}, {Claus},
  {Cohen-Tanugi}, {Conrad}, {Costamante}, {Cutini}, {Dermer}, {de Palma},
  {Donato}, {Silva}, {Drell}, {Dubois}, {Escande}, {Favuzzi}, {Fegan}, {Finke},
  {Focke}, {Fortin}, {Frailis}, {Fukazawa}, {Funk}, {Fusco}, {Gargano},
  {Gasparrini}, {Gehrels}, {Germani}, {Giglietto}, {Giordano}, {Giroletti},
  {Glanzman}, {Godfrey}, {Grenier}, {Guiriec}, {Hadasch}, {Hayashida}, {Hays},
  {Hughes}, {Itoh}, {J{\'o}hannesson}, {Johnson}, {Johnson}, {Kamae},
  {Katagiri}, {Kataoka}, {Kn{\"o}dlseder}, {Kuss}, {Lande}, {Larsson},
  {Latronico}, {Lee}, {Llena Garde}, {Longo}, {Loparco}, {Lott}, {Lovellette},
  {Lubrano}, {Makeev}, {Mazziotta}, {McEnery}, {Mehault}, {Michelson},
  {Mizuno}, {Monte}, {Monzani}, {Morselli}, {Moskalenko}, {Murgia}, {Nakamori},
  {Naumann-Godo}, {Nishino}, {Nolan}, {Norris}, {Nuss}, {Ohsugi}, {Okumura},
  {Omodei}, {Orlando}, {Ormes}, {Ozaki}, {Paneque}, {Panetta}, {Parent},
  {Pelassa}, {Pepe}, {Pesce-Rollins}, {Piron}, {Porter}, {Rain{\`o}}, {Rando},
  {Razzano}, {Reimer}, {Reimer}, {Ritz}, {Roth}, {Sadrozinski}, {Sanchez},
  {Sander}, {Schinzel}, {Sgr{\`o}}, {Siskind}, {Smith}, {Sokolovsky},
  {Spandre}, {Spinelli}, {Strickman}, {Suson}, {Takahashi}, {Tanaka}, {Thayer},
  {Thayer}, {Thompson}, {Tibaldo}, {Torres}, {Tosti}, {Tramacere}, {Uehara},
  {Usher}, {Vandenbroucke}, {Vasileiou}, {Vilchez}, {Vitale}, {Waite},
  {Wallace}, {Wang}, {Winer}, {Wood}, {Yang}, {Ylinen}, {Ziegler}, {Berdyugin},
  {Boettcher}, {Carrami{\~n}ana}, {Carrasco}, {de la Fuente}, {Diltz},
  {Hovatta}, {Kadenius}, {Kovalev}, {L{\"a}hteenm{\"a}ki}, {Lindfors},
  {Marscher}, {Nilsson}, {Pereira}, {Reinthal}, {Roustazadeh}, {Savolainen},
  {Sillanp{\"a}{\"a}}, {Takalo}, \& {Tornikoski}}]{Abdo_2011}
---. 2011{\natexlab{b}}, \apj, 730, 101, \dodoi{10.1088/0004-637X/730/2/101}

\bibitem[{{Abdollahi} {et~al.}(2022){Abdollahi}, {Acero}, {Baldini}, {Ballet},
  {Bastieri}, {Bellazzini}, {Berenji}, {Berretta}, {Bissaldi}, {Blandford},
  {Bloom}, {Bonino}, {Brill}, {Britto}, {Bruel}, {Burnett}, {Buson}, {Cameron},
  {Caputo}, {Caraveo}, {Castro}, {Chaty}, {Cheung}, {Chiaro}, {Cibrario},
  {Ciprini}, {Coronado-Bl{\'a}zquez}, {Crnogorcevic}, {Cutini}, {D'Ammando},
  {De Gaetano}, {Digel}, {Di Lalla}, {Dirirsa}, {Di Venere}, {Dom{\'\i}nguez},
  {Fallah Ramazani}, {Fegan}, {Ferrara}, {Fiori}, {Fleischhack}, {Franckowiak},
  {Fukazawa}, {Funk}, {Fusco}, {Galanti}, {Gammaldi}, {Gargano}, {Garrappa},
  {Gasparrini}, {Giacchino}, {Giglietto}, {Giordano}, {Giroletti}, {Glanzman},
  {Green}, {Grenier}, {Grondin}, {Guillemot}, {Guiriec}, {Gustafsson},
  {Harding}, {Hays}, {Hewitt}, {Horan}, {Hou}, {J{\'o}hannesson}, {Karwin},
  {Kayanoki}, {Kerr}, {Kuss}, {Landriu}, {Larsson}, {Latronico},
  {Lemoine-Goumard}, {Li}, {Liodakis}, {Longo}, {Loparco}, {Lott}, {Lubrano},
  {Maldera}, {Malyshev}, {Manfreda}, {Mart{\'\i}-Devesa}, {Mazziotta}, {Mereu},
  {Meyer}, {Michelson}, {Mirabal}, {Mitthumsiri}, {Mizuno}, {Moiseev},
  {Monzani}, {Morselli}, {Moskalenko}, {Negro}, {Nuss}, {Omodei}, {Orienti},
  {Orlando}, {Paneque}, {Pei}, {Perkins}, {Persic}, {Pesce-Rollins},
  {Petrosian}, {Pillera}, {Poon}, {Porter}, {Principe}, {Rain{\`o}}, {Rando},
  {Rani}, {Razzano}, {Razzaque}, {Reimer}, {Reimer}, {Reposeur},
  {S{\'a}nchez-Conde}, {Saz Parkinson}, {Scotton}, {Serini}, {Sgr{\`o}},
  {Siskind}, {Smith}, {Spandre}, {Spinelli}, {Sueoka}, {Suson}, {Tajima},
  {Tak}, {Thayer}, {Thompson}, {Torres}, {Troja}, {Valverde}, {Wood}, \&
  {Zaharijas}}]{Abdollahi2022}
{Abdollahi}, S., {Acero}, F., {Baldini}, L., {et~al.} 2022, \apjs, 260, 53,
  \dodoi{10.3847/1538-4365/ac6751}

\bibitem[{Ackermann {et~al.}(2011)Ackermann, Ajello, Allafort, Antolini,
  Atwood, Axelsson, Baldini, Ballet, Barbiellini, Bastieri, Bechtol,
  Bellazzini, Berenji, Blandford, Bloom, Bonamente, Borgland, Bottacini,
  Bouvier, Bregeon, Brigida, Bruel, Buehler, Burnett, Buson, Caliandro,
  Cameron, Caraveo, Casandjian, Cavazzuti, Cecchi, Charles, Cheung, Chiang,
  Ciprini, Claus, Cohen-Tanugi, Conrad, Costamante, Cutini, de~Angelis,
  de~Palma, Dermer, Digel, do~Couto~e Silva, Drell, Dubois, Escande, Favuzzi,
  Fegan, Ferrara, Finke, Focke, Fortin, Frailis, Fukazawa, Funk, Fusco,
  Gargano, Gasparrini, Gehrels, Germani, Giebels, Giglietto, Giommi, Giordano,
  Giroletti, Glanzman, Godfrey, Grenier, Grove, Guiriec, Gustafsson, Hadasch,
  Hayashida, Hays, Healey, Horan, Hou, Hughes, Iafrate, Jóhannesson, Johnson,
  Johnson, Kamae, Katagiri, Kataoka, Knödlseder, Kuss, Lande, Larsson,
  Latronico, Longo, Loparco, Lott, Lovellette, Lubrano, Madejski, Mazziotta,
  McConville, McEnery, Michelson, Mitthumsiri, Mizuno, Moiseev, Monte, Monzani,
  Moretti, Morselli, Moskalenko, Murgia, Nakamori, Naumann-Godo, Nolan, Norris,
  Nuss, Ohno, Ohsugi, Okumura, Omodei, Orienti, Orlando, Ormes, Ozaki, Paneque,
  Parent, Pesce-Rollins, Pierbattista, Piranomonte, Piron, Pivato, Porter,
  Rainò, Rando, Razzano, Razzaque, Reimer, Reimer, Ritz, Rochester, Romani,
  Roth, Sanchez, Sbarra, Scargle, Schalk, Sgrò, Shaw, Siskind, Spandre,
  Spinelli, Strong, Suson, Tajima, Takahashi, Takahashi, Tanaka, Thayer,
  Thayer, Thompson, Tibaldo, Tinivella, Torres, Tosti, Troja, Uchiyama,
  Vandenbroucke, Vasileiou, Vianello, Vitale, Waite, Wallace, Wang, Winer,
  Wood, Wood, \& Zimmer}]{Ackermann_2011}
Ackermann, M., Ajello, M., Allafort, A., {et~al.} 2011, The Astrophysical
  Journal, 743, 171, \dodoi{10.1088/0004-637X/743/2/171}

\bibitem[{{Ackermann} {et~al.}(2016){Ackermann}, {Anantua}, {Asano}, {Baldini},
  {Barbiellini}, {Bastieri}, {Becerra Gonzalez}, {Bellazzini}, {Bissaldi},
  {Blandford}, {Bloom}, {Bonino}, {Bottacini}, {Bruel}, {Buehler}, {Caliandro},
  {Cameron}, {Caragiulo}, {Caraveo}, {Cavazzuti}, {Cecchi}, {Cheung}, {Chiang},
  {Chiaro}, {Ciprini}, {Cohen-Tanugi}, {Costanza}, {Cutini}, {D'Ammando}, {de
  Palma}, {Desiante}, {Digel}, {Di Lalla}, {Di Mauro}, {Di Venere}, {Drell},
  {Favuzzi}, {Fegan}, {Ferrara}, {Fukazawa}, {Funk}, {Fusco}, {Gargano},
  {Gasparrini}, {Giglietto}, {Giordano}, {Giroletti}, {Grenier}, {Guillemot},
  {Guiriec}, {Hayashida}, {Hays}, {Horan}, {J{\'o}hannesson}, {Kensei},
  {Kocevski}, {Kuss}, {La Mura}, {Larsson}, {Latronico}, {Li}, {Longo},
  {Loparco}, {Lott}, {Lovellette}, {Lubrano}, {Madejski}, {Magill}, {Maldera},
  {Manfreda}, {Mayer}, {Mazziotta}, {Michelson}, {Mirabal}, {Mizuno},
  {Monzani}, {Morselli}, {Moskalenko}, {Nalewajko}, {Negro}, {Nuss}, {Ohsugi},
  {Orlando}, {Paneque}, {Perkins}, {Pesce-Rollins}, {Piron}, {Pivato},
  {Porter}, {Principe}, {Rando}, {Razzano}, {Razzaque}, {Reimer}, {Scargle},
  {Sgr{\`o}}, {Sikora}, {Simone}, {Siskind}, {Spada}, {Spinelli}, {Stawarz},
  {Thayer}, {Thompson}, {Torres}, {Troja}, {Uchiyama}, {Yuan}, \&
  {Zimmer}}]{Ackermann_16}
{Ackermann}, M., {Anantua}, R., {Asano}, K., {et~al.} 2016, \apjl, 824, L20,
  \dodoi{10.3847/2041-8205/824/2/L20}

\bibitem[{Ackermann {et~al.}(2017)Ackermann, Ajello, Baldini, Ballet,
  Barbiellini, Bastieri, Gonzalez, Bellazzini, Bissaldi, Blandford, Bloom,
  Bonino, Bottacini, Bregeon, Bruel, Buehler, Buson, Cameron, Caragiulo,
  Caraveo, Cavazzuti, Cecchi, Cheung, Chiang, Chiaro, Ciprini, Conrad,
  Costantin, Costanza, Cutini, D’Ammando, de~Palma, Desiante, Digel, Lalla,
  Mauro, Venere, Domínguez, Drell, Favuzzi, Fegan, Ferrara, Finke, Focke,
  Fukazawa, Funk, Fusco, Gargano, Gasparrini, Giglietto, Giordano, Giroletti,
  Green, Grenier, Guillemot, Guiriec, Hartmann, Hays, Horan, Jogler,
  Jóhannesson, Johnson, Kuss, Mura, Larsson, Latronico, Li, Longo, Loparco,
  Lovellette, Lubrano, Magill, Maldera, Manfreda, Marcotulli, Mazziotta,
  Michelson, Mirabal, Mitthumsiri, Mizuno, Monzani, Morselli, Moskalenko,
  Negro, Nuss, Ohsugi, Ojha, Omodei, Orienti, Orlando, Ormes, Paliya, Paneque,
  Perkins, Persic, Pesce-Rollins, Piron, Porter, Principe, Rainò, Rando, Rani,
  Razzano, Razzaque, Reimer, Reimer, Romani, Sgrò, Simone, Siskind, Spada,
  Spandre, Spinelli, Stalin, Stawarz, Suson, Takahashi, Tanaka, Thayer,
  Thompson, Torres, Torresi, Tosti, Troja, Vianello, \& Wood}]{Ackermann_2017}
Ackermann, M., Ajello, M., Baldini, L., {et~al.} 2017, The Astrophysical
  Journal Letters, 837, L5, \dodoi{10.3847/2041-8213/aa5fff}

\bibitem[{Ajello {et~al.}(2020)Ajello, Angioni, Axelsson, Ballet, Barbiellini,
  Bastieri, Gonzalez, Bellazzini, Bissaldi, Bloom, Bonino, Bottacini, Bruel,
  Buson, Cafardo, Cameron, Cavazzuti, Chen, Cheung, Ciprini, Costantin, Cutini,
  D’Ammando, de~la Torre~Luque, de~Menezes, de~Palma, Desai, Lalla, Venere,
  Domínguez, Dirirsa, Ferrara, Finke, Franckowiak, Fukazawa, Funk, Fusco,
  Gargano, Garrappa, Gasparrini, Giglietto, Giordano, Giroletti, Green,
  Grenier, Guiriec, Harita, Hays, Horan, Itoh, Jóhannesson, Kovac’evic’,
  Krauss, Kreter, Kuss, Larsson, Leto, Li, Liodakis, Longo, Loparco, Lott,
  Lovellette, Lubrano, Madejski, Maldera, Manfreda, Martí-Devesa, Massaro,
  Mazziotta, Mereu, Meyer, Migliori, Mirabal, Mizuno, Monzani, Morselli,
  Moskalenko, Negro, Nemmen, Nuss, Ojha, Ojha, Omodei, Orienti, Orlando, Ormes,
  Paliya, Pei, Peña-Herazo, Persic, Pesce-Rollins, Petrov, Piron, Poon,
  Principe, Rainò, Rando, Rani, Razzano, Razzaque, Reimer, Reimer, Schinzel,
  Serini, Sgrò, Siskind, Spandre, Spinelli, Suson, Tachibana, Thompson,
  Torres, Torresi, Troja, Valverde, van Zyl, \& Yassine}]{Ajello_2020}
Ajello, M., Angioni, R., Axelsson, M., {et~al.} 2020, The Astrophysical
  Journal, 892, 105, \dodoi{10.3847/1538-4357/ab791e}

\bibitem[{{Arnaud}(1996)}]{Arnaud_Xspec}
{Arnaud}, K.~A. 1996, in Astronomical Society of the Pacific Conference Series,
  Vol. 101, Astronomical Data Analysis Software and Systems V, ed. G.~H.
  {Jacoby} \& J.~{Barnes}, 17

\bibitem[{{Astropy Collaboration} {et~al.}(2013){Astropy Collaboration},
  {Robitaille}, {Tollerud}, {Greenfield}, {Droettboom}, {Bray}, {Aldcroft},
  {Davis}, {Ginsburg}, {Price-Whelan}, {Kerzendorf}, {Conley}, {Crighton},
  {Barbary}, {Muna}, {Ferguson}, {Grollier}, {Parikh}, {Nair}, {Unther},
  {Deil}, {Woillez}, {Conseil}, {Kramer}, {Turner}, {Singer}, {Fox}, {Weaver},
  {Zabalza}, {Edwards}, {Azalee Bostroem}, {Burke}, {Casey}, {Crawford},
  {Dencheva}, {Ely}, {Jenness}, {Labrie}, {Lim}, {Pierfederici}, {Pontzen},
  {Ptak}, {Refsdal}, {Servillat}, \& {Streicher}}]{2013A&A...558A..33A}
{Astropy Collaboration}, {Robitaille}, T.~P., {Tollerud}, E.~J., {et~al.} 2013,
  \aap, 558, A33, \dodoi{10.1051/0004-6361/201322068}

\bibitem[{{Astropy Collaboration} {et~al.}(2018){Astropy Collaboration},
  {Price-Whelan}, {Sip{\H{o}}cz}, {G{\"u}nther}, {Lim}, {Crawford}, {Conseil},
  {Shupe}, {Craig}, {Dencheva}, {Ginsburg}, {VanderPlas}, {Bradley},
  {P{\'e}rez-Su{\'a}rez}, {de Val-Borro}, {Aldcroft}, {Cruz}, {Robitaille},
  {Tollerud}, {Ardelean}, {Babej}, {Bach}, {Bachetti}, {Bakanov}, {Bamford},
  {Barentsen}, {Barmby}, {Baumbach}, {Berry}, {Biscani}, {Boquien}, {Bostroem},
  {Bouma}, {Brammer}, {Bray}, {Breytenbach}, {Buddelmeijer}, {Burke},
  {Calderone}, {Cano Rodr{\'\i}guez}, {Cara}, {Cardoso}, {Cheedella}, {Copin},
  {Corrales}, {Crichton}, {D'Avella}, {Deil}, {Depagne}, {Dietrich}, {Donath},
  {Droettboom}, {Earl}, {Erben}, {Fabbro}, {Ferreira}, {Finethy}, {Fox},
  {Garrison}, {Gibbons}, {Goldstein}, {Gommers}, {Greco}, {Greenfield},
  {Groener}, {Grollier}, {Hagen}, {Hirst}, {Homeier}, {Horton}, {Hosseinzadeh},
  {Hu}, {Hunkeler}, {Ivezi{\'c}}, {Jain}, {Jenness}, {Kanarek}, {Kendrew},
  {Kern}, {Kerzendorf}, {Khvalko}, {King}, {Kirkby}, {Kulkarni}, {Kumar},
  {Lee}, {Lenz}, {Littlefair}, {Ma}, {Macleod}, {Mastropietro}, {McCully},
  {Montagnac}, {Morris}, {Mueller}, {Mumford}, {Muna}, {Murphy}, {Nelson},
  {Nguyen}, {Ninan}, {N{\"o}the}, {Ogaz}, {Oh}, {Parejko}, {Parley}, {Pascual},
  {Patil}, {Patil}, {Plunkett}, {Prochaska}, {Rastogi}, {Reddy Janga},
  {Sabater}, {Sakurikar}, {Seifert}, {Sherbert}, {Sherwood-Taylor}, {Shih},
  {Sick}, {Silbiger}, {Singanamalla}, {Singer}, {Sladen}, {Sooley},
  {Sornarajah}, {Streicher}, {Teuben}, {Thomas}, {Tremblay}, {Turner},
  {Terr{\'o}n}, {van Kerkwijk}, {de la Vega}, {Watkins}, {Weaver}, {Whitmore},
  {Woillez}, {Zabalza}, \& {Astropy Contributors}}]{2018AJ....156..123A}
{Astropy Collaboration}, {Price-Whelan}, A.~M., {Sip{\H{o}}cz}, B.~M., {et~al.}
  2018, \aj, 156, 123, \dodoi{10.3847/1538-3881/aabc4f}

\bibitem[{Bachev {et~al.}(2023)Bachev, Tripathi, Gupta, Kushwaha, Strigachev,
  Kurtenkov, Nikolov, Boeva, Damljanovic, Vince, Stojanovic, Kishore, Gaur,
  Dhiman, Fan, Kalita, Spassov, \& Semkov}]{Bachev_2023}
Bachev, R., Tripathi, T., Gupta, A.~C., {et~al.} 2023, Monthly Notices of the
  Royal Astronomical Society, 522, 3018, \dodoi{10.1093/mnras/stad1063}

\bibitem[{{Banados} {et~al.}(2024){Banados}, {Momjian}, {Connor}, {Belladitta},
  {Decarli}, {Mazzucchelli}, {Venemans}, {Walter}, {Wang}, {Xie}, {Barth},
  {Eilers}, {Fan}, {Khusanova}, {Schindler}, {Stern}, {Yang}, {Taufik Andika},
  {Carilli}, {Farina}, {Fabian}, {Hennawi}, {Pensabene}, \&
  {Rojas-Ruiz}}]{Banados_2024}
{Banados}, E., {Momjian}, E., {Connor}, T., {et~al.} 2024, arXiv e-prints,
  arXiv:2407.07236, \dodoi{10.48550/arXiv.2407.07236}

\bibitem[{{Blandford} \& {Ostriker}(1978)}]{Blandford_1978}
{Blandford}, R.~D., \& {Ostriker}, J.~P. 1978, \apjl, 221, L29,
  \dodoi{10.1086/182658}

\bibitem[{{B{\l}a{\.z}ejowski} {et~al.}(2000){B{\l}a{\.z}ejowski}, {Sikora},
  {Moderski}, \& {Madejski}}]{Blazejowski_2000}
{B{\l}a{\.z}ejowski}, M., {Sikora}, M., {Moderski}, R., \& {Madejski}, G.~M.
  2000, \apj, 545, 107, \dodoi{10.1086/317791}

\bibitem[{{Bloom} \& {Marscher}(1996)}]{Bloom_Marscher_1996}
{Bloom}, S.~D., \& {Marscher}, A.~P. 1996, \apj, 461, 657,
  \dodoi{10.1086/177092}

\bibitem[{Bloom {et~al.}(1997)Bloom, Bertsch, Hartman, Sreekumar, Thompson,
  Balonek, Beckerman, Davis, Whitman, Miller, Nair, L.~C.~Roberts, Tosti,
  Massaro, Nesci, Maesano, Montagni, Jang, Bock, Dietrich, Herter, Otterbein,
  Pfeiffer, Seitz, \& Wagner}]{Bloom_1997}
Bloom, S.~D., Bertsch, D.~L., Hartman, R.~C., {et~al.} 1997, The Astrophysical
  Journal, 490, L145, \dodoi{10.1086/311035}

\bibitem[{{Bottacini} {et~al.}(2016){Bottacini}, {B{\"o}ttcher}, {Pian}, \&
  {Collmar}}]{botta16}
{Bottacini}, E., {B{\"o}ttcher}, M., {Pian}, E., \& {Collmar}, W. 2016, \apj,
  832, 17, \dodoi{10.3847/0004-637X/832/1/17}

\bibitem[{{B{\"o}ttcher} \& {Bloom}(2000)}]{Bottcher_2000}
{B{\"o}ttcher}, M., \& {Bloom}, S.~D. 2000, \aj, 119, 469,
  \dodoi{10.1086/301201}

\bibitem[{{B{\"o}ttcher} \& {Reimer}(2004)}]{Bottcher_2004}
{B{\"o}ttcher}, M., \& {Reimer}, A. 2004, \apj, 609, 576,
  \dodoi{10.1086/421320}

\bibitem[{{B{\"o}ttcher} {et~al.}(2013){B{\"o}ttcher}, {Reimer}, {Sweeney}, \&
  {Prakash}}]{bottcher13}
{B{\"o}ttcher}, M., {Reimer}, A., {Sweeney}, K., \& {Prakash}, A. 2013, \apj,
  768, 54, \dodoi{10.1088/0004-637X/768/1/54}

\bibitem[{{Burrows} {et~al.}(2005){Burrows}, {Hill}, {Nousek}, {Kennea},
  {Wells}, {Osborne}, {Abbey}, {Beardmore}, {Mukerjee}, {Short}, {Chincarini},
  {Campana}, {Citterio}, {Moretti}, {Pagani}, {Tagliaferri}, {Giommi},
  {Capalbi}, {Tamburelli}, {Angelini}, {Cusumano}, {Br{\"a}uninger}, {Burkert},
  \& {Hartner}}]{Swift_XRT}
{Burrows}, D.~N., {Hill}, J.~E., {Nousek}, J.~A., {et~al.} 2005, \ssr, 120,
  165, \dodoi{10.1007/s11214-005-5097-2}

\bibitem[{{Capetti} {et~al.}(2010){Capetti}, {Raiteri}, \&
  {Buttiglione}}]{Capetti_2010}
{Capetti}, A., {Raiteri}, C.~M., \& {Buttiglione}, S. 2010, \aap, 516, A59,
  \dodoi{10.1051/0004-6361/201014232}

\bibitem[{{Cardelli} {et~al.}(1989){Cardelli}, {Clayton}, \&
  {Mathis}}]{Cardelli_1989}
{Cardelli}, J.~A., {Clayton}, G.~C., \& {Mathis}, J.~S. 1989, \apj, 345, 245,
  \dodoi{10.1086/167900}

\bibitem[{Chiang \& Böttcher(2002)}]{Chiang_2002}
Chiang, J., \& Böttcher, M. 2002, The Astrophysical Journal, 564, 92,
  \dodoi{10.1086/324294}

\bibitem[{Cleary {et~al.}(2007)Cleary, Lawrence, Marshall, Hao, \&
  Meier}]{Cleary_2007}
Cleary, K., Lawrence, C.~R., Marshall, J.~A., Hao, L., \& Meier, D. 2007, The
  Astrophysical Journal, 660, 117, \dodoi{10.1086/511969}

\bibitem[{{Corbett} {et~al.}(1996){Corbett}, {Robinson}, {Axon}, {Hough},
  {Jeffries}, {Thurston}, \& {Young}}]{Corbett_1996}
{Corbett}, E.~A., {Robinson}, A., {Axon}, D.~J., {et~al.} 1996, \mnras, 281,
  737, \dodoi{10.1093/mnras/281.3.737}

\bibitem[{{Dermer} {et~al.}(2009){Dermer}, {Finke}, {Krug}, \&
  {B{\"o}ttcher}}]{Dermer_2009}
{Dermer}, C.~D., {Finke}, J.~D., {Krug}, H., \& {B{\"o}ttcher}, M. 2009, \apj,
  692, 32, \dodoi{10.1088/0004-637X/692/1/32}

\bibitem[{{Di Gesu} {et~al.}(2022){Di Gesu}, {Donnarumma}, {Tavecchio},
  {Agudo}, {Barnounin}, {Cibrario}, {Di Lalla}, {Di Marco}, {Escudero},
  {Errando}, {Jorstad}, {Kim}, {Kouch}, {Liodakis}, {Lindfors}, {Madejski},
  {Marshall}, {Marscher}, {Middei}, {Muleri}, {Myserlis}, {Negro}, {Omodei},
  {Pacciani}, {Paggi}, {Perri}, {Puccetti}, {Antonelli}, {Bachetti}, {Baldini},
  {Baumgartner}, {Bellazzini}, {Bianchi}, {Bongiorno}, {Bonino}, {Brez},
  {Bucciantini}, {Capitanio}, {Castellano}, {Cavazzuti}, {Ciprini}, {Costa},
  {De Rosa}, {Del Monte}, {Doroshenko}, {Dov{\v{c}}iak}, {Ehlert}, {Enoto},
  {Evangelista}, {Fabiani}, {Ferrazzoli}, {Garcia}, {Gunji}, {Hayashida},
  {Heyl}, {Iwakiri}, {Karas}, {Kitaguchi}, {Kolodziejczak}, {Krawczynski}, {La
  Monaca}, {Latronico}, {Maldera}, {Manfreda}, {Marin}, {Marinucci}, {Massaro},
  {Matt}, {Mitsuishi}, {Mizuno}, {Ng}, {O'Dell}, {Oppedisano}, {Papitto},
  {Pavlov}, {Peirson}, {Pesce-Rollins}, {Petrucci}, {Pilia}, {Possenti},
  {Poutanen}, {Ramsey}, {Rankin}, {Ratheesh}, {Romani}, {Sgr{\`o}}, {Slane},
  {Soffitta}, {Spandre}, {Tamagawa}, {Taverna}, {Tawara}, {Tennant}, {Thomas},
  {Tombesi}, {Trois}, {Tsygankov}, {Turolla}, {Vink}, {Weisskopf}, {Wu}, {Xie},
  \& {Zane}}]{Di_Gesu_2022}
{Di Gesu}, L., {Donnarumma}, I., {Tavecchio}, F., {et~al.} 2022, \apjl, 938,
  L7, \dodoi{10.3847/2041-8213/ac913a}

\bibitem[{{Di Gesu} {et~al.}(2023){Di Gesu}, {Marshall}, {Ehlert}, {Kim},
  {Donnarumma}, {Tavecchio}, {Liodakis}, {Kiehlmann}, {Agudo}, {Jorstad},
  {Muleri}, {Marscher}, {Puccetti}, {Middei}, {Perri}, {Pacciani}, {Negro},
  {Romani}, {Di Marco}, {Blinov}, {Bourbah}, {Kontopodis}, {Mandarakas},
  {Romanopoulos}, {Skalidis}, {Vervelaki}, {Casadio}, {Escudero}, {Myserlis},
  {Gurwell}, {Rao}, {Keating}, {Kouch}, {Lindfors}, {Aceituno}, {Bernardos},
  {Bonnoli}, {Casanova}, {Garc{\'\i}a-Comas}, {Ag{\'\i}s-Gonz{\'a}lez},
  {Husillos}, {Marchini}, {Sota}, {Imazawa}, {Sasada}, {Fukazawa}, {Kawabata},
  {Uemura}, {Mizuno}, {Nakaoka}, {Akitaya}, {Savchenko}, {Vasilyev},
  {G{\'o}mez}, {Antonelli}, {Barnouin}, {Bonino}, {Cavazzuti}, {Costamante},
  {Chen}, {Cibrario}, {De Rosa}, {Di Pierro}, {Errando}, {Kaaret}, {Karas},
  {Krawczynski}, {Lisalda}, {Madejski}, {Malacaria}, {Marin}, {Marinucci},
  {Massaro}, {Matt}, {Mitsuishi}, {O'Dell}, {Paggi}, {Peirson}, {Petrucci},
  {Ramsey}, {Tennant}, {Wu}, {Bachetti}, {Baldini}, {Baumgartner},
  {Bellazzini}, {Bianchi}, {Bongiorno}, {Brez}, {Bucciantini}, {Capitanio},
  {Castellano}, {Ciprini}, {Costa}, {Del Monte}, {Di Lalla}, {Doroshenko},
  {Dov{\v{c}}iak}, {Enoto}, {Evangelista}, {Fabiani}, {Ferrazzoli}, {Garcia},
  {Gunji}, {Hayashida}, {Heyl}, {Iwakiri}, {Kislat}, {Kitaguchi},
  {Kolodziejczak}, {La Monaca}, {Latronico}, {Maldera}, {Manfreda}, {Ng},
  {Omodei}, {Oppedisano}, {Papitto}, {Pavlov}, {Pesce-Rollins}, {Pilia},
  {Possenti}, {Poutanen}, {Rankin}, {Ratheesh}, {Roberts}, {Sgr{\`o}}, {Slane},
  {Soffitta}, {Spandre}, {Swartz}, {Tamagawa}, {Taverna}, {Tawara}, {Thomas},
  {Tombesi}, {Trois}, {Tsygankov}, {Turolla}, {Vink}, {Weisskopf}, {Xie}, \&
  {Zane}}]{Di_Gesu_2023NatAs}
{Di Gesu}, L., {Marshall}, H.~L., {Ehlert}, S.~R., {et~al.} 2023, Nature
  Astronomy, 7, 1245, \dodoi{10.1038/s41550-023-02032-7}

\bibitem[{{Dorman} \& {Arnaud}(2001)}]{Dorman_Xspec}
{Dorman}, B., \& {Arnaud}, K.~A. 2001, in Astronomical Society of the Pacific
  Conference Series, Vol. 238, Astronomical Data Analysis Software and Systems
  X, ed. J.~{Harnden}, F.~R., F.~A. {Primini}, \& H.~E. {Payne}, 415

\bibitem[{{Dorman} {et~al.}(2003){Dorman}, {Arnaud}, \& {Gordon}}]{Dorman_2003}
{Dorman}, B., {Arnaud}, K.~A., \& {Gordon}, C.~A. 2003, in AAS/High Energy
  Astrophysics Division, Vol.~7, AAS/High Energy Astrophysics Division \#7,
  22.10

\bibitem[{Ehlert {et~al.}(2023)Ehlert, Liodakis, Middei, Marscher, Tavecchio,
  Agudo, Kouch, Lindfors, Nilsson, Myserlis, Gurwell, Rao, Aceituno, Bonnoli,
  Casanova, Agís-González, Escudero, Husillos, Santos, Sota, Angelakis,
  Kraus, Keating, Antonelli, Bachetti, Baldini, Baumgartner, Bellazzini,
  Bianchi, Bongiorno, Bonino, Brez, Bucciantini, Capitanio, Castellano,
  Cavazzuti, Chen, Ciprini, Costa, Rosa, Monte, Gesu, Lalla, Marco, Donnarumma,
  Doroshenko, Dovčiak, Enoto, Evangelista, Fabiani, Ferrazzoli, Garcia, Gunji,
  Hayashida, Heyl, Iwakiri, Jorstad, Kaaret, Karas, Kislat, Kitaguchi,
  Kolodziejczak, Krawczynski, Monaca, Latronico, Maldera, Manfreda, Marin,
  Marinucci, Marshall, Massaro, Matt, Mitsuishi, Mizuno, Muleri, Negro, Ng,
  O’Dell, Omodei, Oppedisano, Papitto, Pavlov, Peirson, Perri, Pesce-Rollins,
  Petrucci, Pilia, Possenti, Poutanen, Puccetti, Ramsey, Rankin, Ratheesh,
  Roberts, Romani, Sgró, Slane, Soffitta, Spandre, Swartz, Tamagawa, Taverna,
  Tawara, Tennant, Thomas, Tombesi, Trois, Tsygankov, Turolla, Vink, Weisskopf,
  Wu, Xie, \& Zane}]{Ehlert_2023}
Ehlert, S.~R., Liodakis, I., Middei, R., {et~al.} 2023, The Astrophysical
  Journal, 959, 61, \dodoi{10.3847/1538-4357/ad05c4}

\bibitem[{{Evans} {et~al.}(2009){Evans}, {Beardmore}, {Page}, {Osborne},
  {O'Brien}, {Willingale}, {Starling}, {Burrows}, {Godet}, {Vetere}, {Racusin},
  {Goad}, {Wiersema}, {Angelini}, {Capalbi}, {Chincarini}, {Gehrels}, {Kennea},
  {Margutti}, {Morris}, {Mountford}, {Pagani}, {Perri}, {Romano}, \&
  {Tanvir}}]{Evans_2009}
{Evans}, P.~A., {Beardmore}, A.~P., {Page}, K.~L., {et~al.} 2009, \mnras, 397,
  1177, \dodoi{10.1111/j.1365-2966.2009.14913.x}

\bibitem[{{Fermi Science Support Development Team}(2019)}]{Fermitools}
{Fermi Science Support Development Team}. 2019, {Fermitools: Fermi Science
  Tools}, Astrophysics Source Code Library, record ascl:1905.011

\bibitem[{Fossati {et~al.}(1998)Fossati, Maraschi, Celotti, Comastri, \&
  Ghisellini}]{Fossati_1998}
Fossati, G., Maraschi, L., Celotti, A., Comastri, A., \& Ghisellini, G. 1998,
  Monthly Notices of the Royal Astronomical Society, 299, 433,
  \dodoi{10.1046/j.1365-8711.1998.01828.x}

\bibitem[{Fraija {et~al.}(2017)Fraija, Benítez, Hiriart, Sorcia, López,
  Mújica, Cabrera, de~Diego, Rojas-Luis, Salazar-Vázquez, \&
  Galván-Gámez}]{Fraija_2017}
Fraija, N., Benítez, E., Hiriart, D., {et~al.} 2017, The Astrophysical Journal
  Supplement Series, 232, 7, \dodoi{10.3847/1538-4365/aa82cc}

\bibitem[{{Ghisellini} {et~al.}(1985){Ghisellini}, {Maraschi}, \&
  {Treves}}]{Ghisellini_1985}
{Ghisellini}, G., {Maraschi}, L., \& {Treves}, A. 1985, \aap, 146, 204

\bibitem[{{Ghisellini} \& {Tavecchio}(2009)}]{Ghisellini_2009}
{Ghisellini}, G., \& {Tavecchio}, F. 2009, \mnras, 397, 985,
  \dodoi{10.1111/j.1365-2966.2009.15007.x}

\bibitem[{Ghisellini {et~al.}(2011)Ghisellini, Tagliaferri, Foschini,
  Ghirlanda, Tavecchio, Ceca, Haardt, Volonteri, \& Gehrels}]{Ghisellini_2011}
Ghisellini, G., Tagliaferri, G., Foschini, L., {et~al.} 2011, Monthly Notices
  of the Royal Astronomical Society, 411, 901,
  \dodoi{10.1111/j.1365-2966.2010.17723.x}

\bibitem[{{Gurwell} {et~al.}(2023){Gurwell}, {Rao}, \& {SMA
  Team}}]{2023ATel16340....1G}
{Gurwell}, M., {Rao}, R., \& {SMA Team}. 2023, The Astronomer's Telegram,
  16340, 1

\bibitem[{{Harrison} {et~al.}(2013){Harrison}, {Craig}, {Christensen},
  {Hailey}, {Zhang}, {Boggs}, {Stern}, {Cook}, {Forster}, {Giommi},
  {Grefenstette}, {Kim}, {Kitaguchi}, {Koglin}, {Madsen}, {Mao}, {Miyasaka},
  {Mori}, {Perri}, {Pivovaroff}, {Puccetti}, {Rana}, {Westergaard}, {Willis},
  {Zoglauer}, {An}, {Bachetti}, {Barri{\`e}re}, {Bellm}, {Bhalerao},
  {Brejnholt}, {Fuerst}, {Liebe}, {Markwardt}, {Nynka}, {Vogel}, {Walton},
  {Wik}, {Alexander}, {Cominsky}, {Hornschemeier}, {Hornstrup}, {Kaspi},
  {Madejski}, {Matt}, {Molendi}, {Smith}, {Tomsick}, {Ajello}, {Ballantyne},
  {Balokovi{\'c}}, {Barret}, {Bauer}, {Blandford}, {Brandt}, {Brenneman},
  {Chiang}, {Chakrabarty}, {Chenevez}, {Comastri}, {Dufour}, {Elvis}, {Fabian},
  {Farrah}, {Fryer}, {Gotthelf}, {Grindlay}, {Helfand}, {Krivonos}, {Meier},
  {Miller}, {Natalucci}, {Ogle}, {Ofek}, {Ptak}, {Reynolds}, {Rigby},
  {Tagliaferri}, {Thorsett}, {Treister}, \& {Urry}}]{Harrison_NuSTAR}
{Harrison}, F.~A., {Craig}, W.~W., {Christensen}, F.~E., {et~al.} 2013, \apj,
  770, 103, \dodoi{10.1088/0004-637X/770/2/103}

\bibitem[{{Hawley} \& {Miller}(1977)}]{Hawley_1977}
{Hawley}, S.~A., \& {Miller}, J.~S. 1977, \apj, 212, 94, \dodoi{10.1086/155023}

\bibitem[{Ho {et~al.}(2004)Ho, Moran, \& Lo}]{Ho_2004}
Ho, P. T.~P., Moran, J.~M., \& Lo, K.~Y. 2004, The Astrophysical Journal, 616,
  L1, \dodoi{10.1086/423245}

\bibitem[{Jorstad {et~al.}(2007)Jorstad, Marscher, Stevens, Smith, Forster,
  Gear, Cawthorne, Lister, Stirling, Gómez, Greaves, \& Robson}]{Jorstad_2007}
Jorstad, S.~G., Marscher, A.~P., Stevens, J.~A., {et~al.} 2007, The
  Astronomical Journal, 134, 799, \dodoi{10.1086/519996}

\bibitem[{{Jorstad} {et~al.}(2022){Jorstad}, {Marscher}, {Raiteri}, {Villata},
  {Weaver}, {Zhang}, {Dong}, {G{\'o}mez}, {Perel}, {Savchenko}, {Larionov},
  {Carosati}, {Chen}, {Kurtanidze}, {Marchini}, {Matsumoto}, {Mortari},
  {Aceti}, {Acosta-Pulido}, {Andreeva}, {Apolonio}, {Arena}, {Arkharov},
  {Bachev}, {Banfi}, {Bonnoli}, {Borman}, {Bozhilov}, {Carnerero},
  {Damljanovic}, {Ehgamberdiev}, {Els{\"a}sser}, {Frasca}, {Gabellini},
  {Grishina}, {Gupta}, {Hagen-Thorn}, {Hallum}, {Hart}, {Hasuda}, {Hemrich},
  {Hsiao}, {Ibryamov}, {Irsmambetova}, {Ivanov}, {Joner}, {Kimeridze},
  {Klimanov}, {Kn{\"o}tt}, {Kopatskaya}, {Kurtanidze}, {Kurtenkov}, {Kuutma},
  {Larionova}, {Leonini}, {Lin}, {Lorey}, {Mannheim}, {Marino}, {Minev},
  {Mirzaqulov}, {Morozova}, {Nikiforova}, {Nikolashvili}, {Ovcharov}, {Papini},
  {Pursimo}, {Rahimov}, {Reinhart}, {Sakamoto}, {Salvaggio}, {Semkov},
  {Shakhovskoy}, {Sigua}, {Steineke}, {Stojanovic}, {Strigachev}, {Troitskaya},
  {Troitskiy}, {Tsai}, {Valcheva}, {Vasilyev}, {Vince}, {Waller}, {Zaharieva},
  \& {Chatterjee}}]{Jorstad_2022_nature}
{Jorstad}, S.~G., {Marscher}, A.~P., {Raiteri}, C.~M., {et~al.} 2022, \nat,
  609, 265, \dodoi{10.1038/s41586-022-05038-9}

\bibitem[{Kaspi {et~al.}(2007)Kaspi, Brandt, Maoz, Netzer, Schneider, \&
  Shemmer}]{Kaspi_2007}
Kaspi, S., Brandt, W.~N., Maoz, D., {et~al.} 2007, The Astrophysical Journal,
  659, 997, \dodoi{10.1086/512094}

\bibitem[{{Kirk} {et~al.}(1998){Kirk}, {Rieger}, \& {Mastichiadis}}]{Kirk_1998}
{Kirk}, J.~G., {Rieger}, F.~M., \& {Mastichiadis}, A. 1998, \aap, 333, 452,
  \dodoi{10.48550/arXiv.astro-ph/9801265}

\bibitem[{{Kouch} {et~al.}(2024){Kouch}, {Liodakis}, {Middei}, {Kim},
  {Tavecchio}, {Marscher}, {Marshall}, {Ehlert}, {Di Gesu}, {Jorstad}, {Agudo},
  {Madejski}, {Romani}, {Errando}, {Lindfors}, {Nilsson}, {Toppari}, {Potter},
  {Imazawa}, {Sasada}, {Fukazawa}, {Kawabata}, {Uemura}, {Mizuno}, {Nakaoka},
  {Akitaya}, {McCall}, {Jermak}, {Steele}, {Myserlis}, {Gurwell}, {Keating},
  {Rao}, {Kang}, {Lee}, {Kim}, {Cheong}, {Jeong}, {Angelakis}, {Kraus},
  {Jos{\'e} Aceituno}, {Bonnoli}, {Casanova}, {Escudero},
  {Ag{\'\i}s-Gonz{\'a}lez}, {Husillos}, {Morcuende}, {Otero-Santos}, {Sota},
  {Bachev}, {Antonelli}, {Bachetti}, {Baldini}, {Baumgartner}, {Bellazzini},
  {Bianchi}, {Bongiorno}, {Bonino}, {Brez}, {Bucciantini}, {Capitanio},
  {Castellano}, {Cavazzuti}, {Chen}, {Ciprini}, {Costa}, {De Rosa}, {Del
  Monte}, {Di Lalla}, {Di Marco}, {Donnarumma}, {Doroshenko}, {Dov{\v{c}}iak},
  {Enoto}, {Evangelista}, {Fabiani}, {Ferrazzoli}, {Garcia}, {Gunji},
  {Hayashida}, {Heyl}, {Iwakiri}, {Kaaret}, {Karas}, {Kislat}, {Kitaguchi},
  {Kolodziejczak}, {Krawczynski}, {La Monaca}, {Latronico}, {Maldera},
  {Manfreda}, {Marin}, {Marinucci}, {Massaro}, {Matt}, {Mitsuishi}, {Muleri},
  {Negro}, {Ng}, {O'Dell}, {Omodei}, {Oppedisano}, {Papitto}, {Pavlov},
  {Peirson}, {Perri}, {Pesce-Rollins}, {Petrucci}, {Pilia}, {Possenti},
  {Poutanen}, {Puccetti}, {Ramsey}, {Rankin}, {Ratheesh}, {Roberts},
  {Sgr{\`o}}, {Slane}, {Soffitta}, {Spandre}, {Swartz}, {Tamagawa}, {Taverna},
  {Tawara}, {Tennant}, {Thomas}, {Tombesi}, {Trois}, {Tsygankov}, {Turolla},
  {Vink}, {Weisskopf}, {Wu}, {Xie}, \& {Zane}}]{Kouch_2024}
{Kouch}, P.~M., {Liodakis}, I., {Middei}, R., {et~al.} 2024, arXiv e-prints,
  arXiv:2406.01693, \dodoi{10.48550/arXiv.2406.01693}

\bibitem[{Krawczynski(2011)}]{Krawczynski_2012}
Krawczynski, H. 2011, The Astrophysical Journal, 744, 30,
  \dodoi{10.1088/0004-637X/744/1/30}

\bibitem[{{Kwan} \& {Krolik}(1981)}]{Kwan_1981}
{Kwan}, J., \& {Krolik}, J.~H. 1981, \apj, 250, 478, \dodoi{10.1086/159395}

\bibitem[{{Larionov} {et~al.}(2010){Larionov}, {Villata}, \&
  {Raiteri}}]{Larionov_2010}
{Larionov}, V.~M., {Villata}, M., \& {Raiteri}, C.~M. 2010, \aap, 510, A93,
  \dodoi{10.1051/0004-6361/200913536}

\bibitem[{Liodakis {et~al.}(2019)Liodakis, Peirson, \& Romani}]{Liodakis_2019}
Liodakis, I., Peirson, A.~L., \& Romani, R.~W. 2019, The Astrophysical Journal,
  880, 29, \dodoi{10.3847/1538-4357/ab2719}

\bibitem[{{Liodakis} {et~al.}(2022){Liodakis}, {Marscher}, {Agudo},
  {Berdyugin}, {Bernardos}, {Bonnoli}, {Borman}, {Casadio}, {Casanova},
  {Cavazzuti}, {Rodriguez Cavero}, {Di Gesu}, {Di Lalla}, {Donnarumma},
  {Ehlert}, {Errando}, {Escudero}, {Garc{\'\i}a-Comas},
  {Ag{\'\i}s-Gonz{\'a}lez}, {Husillos}, {Jormanainen}, {Jorstad}, {Kagitani},
  {Kopatskaya}, {Kravtsov}, {Krawczynski}, {Lindfors}, {Larionova}, {Madejski},
  {Marin}, {Marchini}, {Marshall}, {Morozova}, {Massaro}, {Masiero}, {Mawet},
  {Middei}, {Millar-Blanchaer}, {Myserlis}, {Negro}, {Nilsson}, {O'Dell},
  {Omodei}, {Pacciani}, {Paggi}, {Panopoulou}, {Peirson}, {Perri}, {Petrucci},
  {Poutanen}, {Puccetti}, {Romani}, {Sakanoi}, {Savchenko}, {Sota},
  {Tavecchio}, {Tinyanont}, {Vasilyev}, {Weaver}, {Zhovtan}, {Antonelli},
  {Bachetti}, {Baldini}, {Baumgartner}, {Bellazzini}, {Bianchi}, {Bongiorno},
  {Bonino}, {Brez}, {Bucciantini}, {Capitanio}, {Castellano}, {Ciprini},
  {Costa}, {De Rosa}, {Del Monte}, {Di Marco}, {Doroshenko}, {Dov{\v{c}}iak},
  {Enoto}, {Evangelista}, {Fabiani}, {Ferrazzoli}, {Garcia}, {Gunji},
  {Hayashida}, {Heyl}, {Iwakiri}, {Karas}, {Kitaguchi}, {Kolodziejczak}, {La
  Monaca}, {Latronico}, {Maldera}, {Manfreda}, {Marinucci}, {Matt},
  {Mitsuishi}, {Mizuno}, {Muleri}, {Ng}, {Oppedisano}, {Papitto}, {Pavlov},
  {Pesce-Rollins}, {Pilia}, {Possenti}, {Ramsey}, {Rankin}, {Ratheesh},
  {Sgr{\'o}}, {Slane}, {Soffitta}, {Spandre}, {Tamagawa}, {Taverna}, {Tawara},
  {Tennant}, {Thomas}, {Tombesi}, {Trois}, {Tsygankov}, {Turolla}, {Vink},
  {Weisskopf}, {Wu}, {Xie}, \& {Zane}}]{Liodakis_2022Nature}
{Liodakis}, I., {Marscher}, A.~P., {Agudo}, I., {et~al.} 2022, \nat, 611, 677,
  \dodoi{10.1038/s41586-022-05338-0}

\bibitem[{{Lisakov} \& {Kovalev}(2015)}]{2015IAUS..313...39L}
{Lisakov}, M.~M., \& {Kovalev}, Y.~Y. 2015, in IAU Symposium, Vol. 313,
  Extragalactic Jets from Every Angle, ed. F.~{Massaro}, C.~C. {Cheung},
  E.~{Lopez}, \& A.~{Siemiginowska}, 39--42, \dodoi{10.1017/S1743921315001830}

\bibitem[{{Lister} {et~al.}(2018){Lister}, {Aller}, {Aller}, {Hodge}, {Homan},
  {Kovalev}, {Pushkarev}, \& {Savolainen}}]{Lister_mojave}
{Lister}, M.~L., {Aller}, M.~F., {Aller}, H.~D., {et~al.} 2018, \apjs, 234, 12,
  \dodoi{10.3847/1538-4365/aa9c44}

\bibitem[{Madejski {et~al.}(1999)Madejski, Sikora, Jaffe, BŁażejowski,
  Jahoda, \& Moderski}]{Madejski_1999}
Madejski, G.~M., Sikora, M., Jaffe, T., {et~al.} 1999, The Astrophysical
  Journal, 521, 145, \dodoi{10.1086/307524}

\bibitem[{{MAGIC Collaboration} {et~al.}(2019){MAGIC Collaboration}, {Acciari},
  {Ansoldi}, {Antonelli}, {Arbet Engels}, {Baack}, {Babi{\'c}}, {Banerjee},
  {Bangale}, {Barres de Almeida}, {Barrio}, {Becerra Gonz{\'a}lez}, {Bednarek},
  {Bernardini}, {Berti}, {Besenrieder}, {Bhattacharyya}, {Bigongiari},
  {Biland}, {Blanch}, {Bonnoli}, {Carosi}, {Ceribella}, {Cikota}, {Colak},
  {Colin}, {Colombo}, {Contreras}, {Cortina}, {Covino}, {D'Elia}, {da Vela},
  {Dazzi}, {de Angelis}, {de Lotto}, {Delfino}, {Delgado}, {di Pierro}, {Do
  Souto Espi{\~n}era}, {Dom{\'\i}nguez}, {Dominis Prester}, {Dorner}, {Doro},
  {Einecke}, {Elsaesser}, {Fallah Ramazani}, {Fattorini},
  {Fern{\'a}ndez-Barral}, {Ferrara}, {Fidalgo}, {Foffano}, {Fonseca}, {Font},
  {Fruck}, {Galindo}, {Gallozzi}, {Garc{\'\i}a L{\'o}pez}, {Garczarczyk},
  {Gaug}, {Giammaria}, {Godinovi{\'c}}, {Guberman}, {Hadasch}, {Hahn},
  {Hassan}, {Herrera}, {Hoang}, {Hrupec}, {Inoue}, {Ishio}, {Iwamura}, {Kubo},
  {Kushida}, {Kuve{\v{z}}di{\'c}}, {Lamastra}, {Lelas}, {Leone}, {Lindfors},
  {Lombardi}, {Longo}, {L{\'o}pez}, {L{\'o}pez-Oramas}, {Maggio}, {Majumdar},
  {Makariev}, {Maneva}, {Manganaro}, {Mannheim}, {Maraschi}, {Mariotti},
  {Mart{\'\i}nez}, {Masuda}, {Mazin}, {Minev}, {Miranda}, {Mirzoyan}, {Molina},
  {Moralejo}, {Moreno}, {Moretti}, {Munar-Adrover}, {Neustroev}, {Niedzwiecki},
  {Nievas Rosillo}, {Nigro}, {Nilsson}, {Ninci}, {Nishijima}, {Noda},
  {Nogu{\'e}s}, {N{\"o}the}, {Paiano}, {Palacio}, {Paneque}, {Paoletti},
  {Paredes}, {Pedaletti}, {Pe{\~n}il}, {Peresano}, {Persic}, {Prada Moroni},
  {Prandini}, {Puljak}, {Garcia}, {Rhode}, {Rib{\'o}}, {Rico}, {Righi},
  {Rugliancich}, {Saha}, {Saito}, {Satalecka}, {Schweizer}, {Sitarek},
  {{\v{S}}nidari{\'c}}, {Sobczynska}, {Somero}, {Stamerra}, {Strzys},
  {Suri{\'c}}, {Tavecchio}, {Temnikov}, {Terzi{\'c}}, {Teshima},
  {Torres-Alb{\`a}}, {Tsujimoto}, {van Scherpenberg}, {Vanzo}, {Vazquez
  Acosta}, {Vovk}, {Will}, {Zari{\'c}}, {D'Ammando}, {Hada}, {Jorstad},
  {Marscher}, {Mobeen}, {Hovatta}, {Larionov}, {Borman}, {Grishina},
  {Kopatskaya}, {Morozova}, {Nikiforova}, {L{\"a}hteenm{\"a}ki}, {Tornikoski},
  \& {Agudo}}]{MAGIC_2019}
{MAGIC Collaboration}, {Acciari}, V.~A., {Ansoldi}, S., {et~al.} 2019, \aap,
  623, A175, \dodoi{10.1051/0004-6361/201834010}

\bibitem[{{Maraschi} {et~al.}(1992){Maraschi}, {Ghisellini}, \&
  {Celotti}}]{Maraschi_1992}
{Maraschi}, L., {Ghisellini}, G., \& {Celotti}, A. 1992, \apjl, 397, L5,
  \dodoi{10.1086/186531}

\bibitem[{{Marscher}(2023)}]{Marscher_book_2023}
{Marscher}, A. 2023, in AAS/High Energy Astrophysics Division, Vol.~20,
  AAS/High Energy Astrophysics Division, 109.01

\bibitem[{Marscher(2013)}]{Marscher_2014}
Marscher, A.~P. 2013, The Astrophysical Journal, 780, 87,
  \dodoi{10.1088/0004-637X/780/1/87}

\bibitem[{{Marscher} {et~al.}(2024){Marscher}, {Di Gesu}, {Jorstad}, {Kim},
  {Liodakis}, {Middei}, \& {Tavecchio}}]{Marscher_2024}
{Marscher}, A.~P., {Di Gesu}, L., {Jorstad}, S.~G., {et~al.} 2024, Galaxies,
  12, 50, \dodoi{10.3390/galaxies12040050}

\bibitem[{{Marscher} {et~al.}(2008){Marscher}, {Jorstad}, {D'Arcangelo},
  {Smith}, {Williams}, {Larionov}, {Oh}, {Olmstead}, {Aller}, {Aller},
  {McHardy}, {L{\"a}hteenm{\"a}ki}, {Tornikoski}, {Valtaoja}, {Hagen-Thorn},
  {Kopatskaya}, {Gear}, {Tosti}, {Kurtanidze}, {Nikolashvili}, {Sigua},
  {Miller}, \& {Ryle}}]{Marscher_2008_Nature}
{Marscher}, A.~P., {Jorstad}, S.~G., {D'Arcangelo}, F.~D., {et~al.} 2008, \nat,
  452, 966, \dodoi{10.1038/nature06895}

\bibitem[{{Middei} {et~al.}(2023){Middei}, {Liodakis}, {Perri}, {Puccetti},
  {Cavazzuti}, {Di Gesu}, {Ehlert}, {Madejski}, {Marscher}, {Marshall},
  {Muleri}, {Negro}, {Jorstad}, {Ag{\'\i}s-Gonz{\'a}lez}, {Agudo}, {Bonnoli},
  {Bernardos}, {Casanova}, {Garc{\'\i}a-Comas}, {Husillos}, {Marchini}, {Sota},
  {Kouch}, {Lindfors}, {Borman}, {Kopatskaya}, {Larionova}, {Morozova},
  {Savchenko}, {Vasilyev}, {Zhovtan}, {Casadio}, {Escudero}, {Myserlis},
  {Hales}, {Kameno}, {Kneissl}, {Messias}, {Nagai}, {Blinov}, {Bourbah},
  {Kiehlmann}, {Kontopodis}, {Mandarakas}, {Romanopoulos}, {Skalidis},
  {Vervelaki}, {Masiero}, {Mawet}, {Millar-Blanchaer}, {Panopoulou},
  {Tinyanont}, {Berdyugin}, {Kagitani}, {Kravtsov}, {Sakanoi}, {Imazawa},
  {Sasada}, {Fukazawa}, {Kawabata}, {Uemura}, {Mizuno}, {Nakaoka}, {Akitaya},
  {Gurwell}, {Rao}, {Di Lalla}, {Cibrario}, {Donnarumma}, {Kim}, {Omodei},
  {Pacciani}, {Poutanen}, {Tavecchio}, {Antonelli}, {Bachetti}, {Baldini},
  {Baumgartner}, {Bellazzini}, {Bianchi}, {Bongiorno}, {Bonino}, {Brez},
  {Bucciantini}, {Capitanio}, {Castellano}, {Ciprini}, {Costa}, {De Rosa}, {Del
  Monte}, {Di Marco}, {Doroshenko}, {Dov{\v{c}}iak}, {Enoto}, {Evangelista},
  {Fabiani}, {Ferrazzoli}, {Garcia}, {Gunji}, {Hayashida}, {Heyl}, {Iwakiri},
  {Karas}, {Kitaguchi}, {Kolodziejczak}, {Krawczynski}, {La Monaca},
  {Latronico}, {Maldera}, {Manfreda}, {Marin}, {Marinucci}, {Massaro}, {Matt},
  {Mitsuishi}, {Ng}, {O'Dell}, {Oppedisano}, {Papitto}, {Pavlov}, {Peirson},
  {Pesce-Rollins}, {Petrucci}, {Pilia}, {Possenti}, {Ramsey}, {Rankin},
  {Ratheesh}, {Romani}, {Sgr{\'o}}, {Slane}, {Soffitta}, {Spandre}, {Tamagawa},
  {Taverna}, {Tawara}, {Tennant}, {Thomas}, {Tombesi}, {Trois}, {Tsygankov},
  {Turolla}, {Vink}, {Weisskopf}, {Wu}, {Xie}, \& {Zane}}]{Middei_2023}
{Middei}, R., {Liodakis}, I., {Perri}, M., {et~al.} 2023, \apjl, 942, L10,
  \dodoi{10.3847/2041-8213/aca281}

\bibitem[{{M{\"u}cke} \& {Protheroe}(2001)}]{mucke01}
{M{\"u}cke}, A., \& {Protheroe}, R.~J. 2001, Astroparticle Physics, 15, 121,
  \dodoi{10.1016/S0927-6505(00)00141-9}

\bibitem[{{Nilsson} {et~al.}(2018){Nilsson}, {Lindfors}, {Takalo}, {Reinthal},
  {Berdyugin}, {Sillanp{\"a}{\"a}}, {Ciprini}, {Halkola}, {Hein{\"a}m{\"a}ki},
  {Hovatta}, {Kadenius}, {Nurmi}, {Ostorero}, {Pasanen}, {Rekola}, {Saarinen},
  {Sainio}, {Tuominen}, {Villforth}, {Vornanen}, \& {Zaprudin}}]{Nilsson_2018}
{Nilsson}, K., {Lindfors}, E., {Takalo}, L.~O., {et~al.} 2018, \aap, 620, A185,
  \dodoi{10.1051/0004-6361/201833621}

\bibitem[{{Paliya} {et~al.}(2021){Paliya}, {Dominguez}, {Ajello},
  {Olmo-Garcia}, \& {Hartmann}}]{Paliya_VizieR_2021}
{Paliya}, V.~S., {Dominguez}, A., {Ajello}, M., {Olmo-Garcia}, A., \&
  {Hartmann}, D. 2021, {VizieR Online Data Catalog: Optical spectroscopy of
  Fermi blazars (Paliya+, 2021)}, VizieR On-line Data Catalog: J/ApJS/253/46.
  Originally published in: 2021ApJS..253...46P

\bibitem[{Peirson \& Romani(2019)}]{Peirson_2019}
Peirson, A.~L., \& Romani, R.~W. 2019, The Astrophysical Journal, 885, 76,
  \dodoi{10.3847/1538-4357/ab46b1}

\bibitem[{{Peirson} {et~al.}(2023){Peirson}, {Negro}, {Liodakis}, {Middei},
  {Kim}, {Marscher}, {Marshall}, {Pacciani}, {Romani}, {Wu}, {Di Marco}, {Di
  Lalla}, {Omodei}, {Jorstad}, {Agudo}, {Kouch}, {Lindfors}, {Aceituno},
  {Bernardos}, {Bonnoli}, {Casanova}, {Garc{\'\i}a-Comas},
  {Ag{\'\i}s-Gonz{\'a}lez}, {Husillos}, {Marchini}, {Sota}, {Casadio},
  {Escudero}, {Myserlis}, {Sievers}, {Gurwell}, {Rao}, {Imazawa}, {Sasada},
  {Fukazawa}, {Kawabata}, {Uemura}, {Mizuno}, {Nakaoka}, {Akitaya}, {Cheong},
  {Jeong}, {Kang}, {Kim}, {Lee}, {Angelakis}, {Kraus}, {Cibrario},
  {Donnarumma}, {Poutanen}, {Tavecchio}, {Antonelli}, {Bachetti}, {Baldini},
  {Baumgartner}, {Bellazzini}, {Bianchi}, {Bongiorno}, {Bonino}, {Brez},
  {Bucciantini}, {Capitanio}, {Castellano}, {Cavazzuti}, {Chen}, {Ciprini},
  {Costa}, {De Rosa}, {Del Monte}, {Di Gesu}, {Doroshenko}, {Dov{\v{c}}iak},
  {Ehlert}, {Enoto}, {Evangelista}, {Fabiani}, {Ferrazzoli}, {Garcia}, {Gunji},
  {Hayashida}, {Heyl}, {Iwakiri}, {Kaaret}, {Karas}, {Kitaguchi},
  {Kolodziejczak}, {Krawczynski}, {La Monaca}, {Latronico}, {Madejski},
  {Maldera}, {Manfreda}, {Marin}, {Marinucci}, {Massaro}, {Matt}, {Mitsuishi},
  {Muleri}, {Ng}, {O'Dell}, {Oppedisano}, {Papitto}, {Pavlov}, {Perri},
  {Pesce-Rollins}, {Petrucci}, {Pilia}, {Possenti}, {Puccetti}, {Ramsey},
  {Rankin}, {Ratheesh}, {Roberts}, {Sgr{\'o}}, {Slane}, {Soffitta}, {Spandre},
  {Swartz}, {Tamagawa}, {Taverna}, {Tawara}, {Tennant}, {Thomas}, {Tombesi},
  {Trois}, {Tsygankov}, {Turolla}, {Vink}, {Weisskopf}, {Xie}, \&
  {Zane}}]{Perison_2023}
{Peirson}, A.~L., {Negro}, M., {Liodakis}, I., {et~al.} 2023, \apjl, 948, L25,
  \dodoi{10.3847/2041-8213/acd242}

\bibitem[{{Potter} \& {Cotter}(2012)}]{Potter_2012}
{Potter}, W.~J., \& {Cotter}, G. 2012, \mnras, 423, 756,
  \dodoi{10.1111/j.1365-2966.2012.20918.x}

\bibitem[{{Pushkarev} {et~al.}(2009){Pushkarev}, {Kovalev}, {Lister}, \&
  {Savolainen}}]{Pushkarev_2009}
{Pushkarev}, A.~B., {Kovalev}, Y.~Y., {Lister}, M.~L., \& {Savolainen}, T.
  2009, \aap, 507, L33, \dodoi{10.1051/0004-6361/200913422}

\bibitem[{{Pushkarev} {et~al.}(2017){Pushkarev}, {Kovalev}, {Lister}, \&
  {Savolainen}}]{Pushkarev_2017}
---. 2017, \mnras, 468, 4992, \dodoi{10.1093/mnras/stx854}

\bibitem[{{Raiteri} {et~al.}(2009){Raiteri}, {Villata}, {Capetti}, {Aller},
  {Bach}, {Calcidese}, {Gurwell}, {Larionov}, {Ohlert}, {Nilsson},
  {Strigachev}, {Agudo}, {Aller}, {Bachev}, {Ben{\'\i}tez}, {Berdyugin},
  {B{\"o}ttcher}, {Buemi}, {Buttiglione}, {Carosati}, {Charlot}, {Chen},
  {Dultzin}, {Forn{\'e}}, {Fuhrmann}, {G{\'o}mez}, {Gupta}, {Heidt}, {Hiriart},
  {Hsiao}, {Jel{\'\i}nek}, {Jorstad}, {Kimeridze}, {Konstantinova},
  {Kopatskaya}, {Kostov}, {Kurtanidze}, {L{\"a}hteenm{\"a}ki}, {Lanteri},
  {Larionova}, {Leto}, {Latev}, {Le Campion}, {Lee}, {Ligustri}, {Lindfors},
  {Marscher}, {Mihov}, {Nikolashvili}, {Nikolov}, {Ovcharov}, {Principe},
  {Pursimo}, {Ragozzine}, {Robb}, {Ros}, {Sadun}, {Sagar}, {Semkov}, {Sigua},
  {Smart}, {Sorcia}, {Takalo}, {Tornikoski}, {Trigilio}, {Uckert}, {Umana},
  {Valcheva}, \& {Volvach}}]{Raiteri_2009}
{Raiteri}, C.~M., {Villata}, M., {Capetti}, A., {et~al.} 2009, \aap, 507, 769,
  \dodoi{10.1051/0004-6361/200912953}

\bibitem[{Raiteri {et~al.}(2013)Raiteri, Villata, D'Ammando, Larionov, Gurwell,
  Mirzaqulov, Smith, Acosta-Pulido, Agudo, Arévalo, Bachev, Benítez,
  Berdyugin, Blinov, Borman, Böttcher, Bozhilov, Carnerero, Carosati, Casadio,
  Chen, Doroshenko, Efimov, Efimova, Ehgamberdiev, Gómez, González-Morales,
  Hiriart, Ibryamov, Jadhav, Jorstad, Joshi, Kadenius, Klimanov, Kohli,
  Konstantinova, Kopatskaya, Koptelova, Kimeridze, Kurtanidze, Larionova,
  Larionova, Ligustri, Lindfors, Marscher, McBreen, McHardy, Metodieva, Molina,
  Morozova, Nazarov, Nikolashvili, Nilsson, Okhmat, Ovcharov, Panwar, Pasanen,
  Peneva, Phipps, Pulatova, Reinthal, Ros, Sadun, Schwartz, Semkov, Sergeev,
  Sigua, Sillanpää, Smith, Stoyanov, Strigachev, Takalo, Taylor, Thum,
  Troitsky, Valcheva, Wehrle, \& Wiesemeyer}]{Raiteri_2013}
Raiteri, C.~M., Villata, M., D'Ammando, F., {et~al.} 2013, Monthly Notices of
  the Royal Astronomical Society, 436, 1530, \dodoi{10.1093/mnras/stt1672}

\bibitem[{Rajguru {et~al.}(2024)Rajguru, Marcotulli, Ajello, \&
  Tramacere}]{Rajguru_2024}
Rajguru, G., Marcotulli, L., Ajello, M., \& Tramacere, A. 2024, The
  Astrophysical Journal, 965, 112, \dodoi{10.3847/1538-4357/ad3236}

\bibitem[{{Roming} {et~al.}(2005){Roming}, {Kennedy}, {Mason}, {Nousek}, {Ahr},
  {Bingham}, {Broos}, {Carter}, {Hancock}, {Huckle}, {Hunsberger}, {Kawakami},
  {Killough}, {Koch}, {McLelland}, {Smith}, {Smith}, {Soto}, {Boyd},
  {Breeveld}, {Holland}, {Ivanushkina}, {Pryzby}, {Still}, \&
  {Stock}}]{Roming_2005_UVOT}
{Roming}, P. W.~A., {Kennedy}, T.~E., {Mason}, K.~O., {et~al.} 2005, \ssr, 120,
  95, \dodoi{10.1007/s11214-005-5095-4}

\bibitem[{{Sahakyan} \& {Giommi}(2022)}]{Sahakyan_2022}
{Sahakyan}, N., \& {Giommi}, P. 2022, \mnras, 513, 4645,
  \dodoi{10.1093/mnras/stac1011}

\bibitem[{Sahakyan {et~al.}(2020)Sahakyan, Israyelyan, Harutyunyan,
  Khachatryan, \& Gasparyan}]{Sahakyan_2020}
Sahakyan, N., Israyelyan, D., Harutyunyan, G., Khachatryan, M., \& Gasparyan,
  S. 2020, Monthly Notices of the Royal Astronomical Society, 498, 2594,
  \dodoi{10.1093/mnras/staa2477}

\bibitem[{Sasada {et~al.}(2014)Sasada, Uemura, Fukazawa, Yasuda, Itoh,
  Sakimoto, Ikejiri, Yoshida, Kawabata, Akitaya, Ohsugi, Yamanaka, Komatsu,
  Miyamoto, Nagae, Nakaya, Tanaka, Sato, \& Kino}]{Sasada_2014}
Sasada, M., Uemura, M., Fukazawa, Y., {et~al.} 2014, The Astrophysical Journal,
  784, 141, \dodoi{10.1088/0004-637X/784/2/141}

\bibitem[{{Schlafly} \& {Finkbeiner}(2011)}]{Schlafly_Finkbeiner_2011}
{Schlafly}, E.~F., \& {Finkbeiner}, D.~P. 2011, \apj, 737, 103,
  \dodoi{10.1088/0004-637X/737/2/103}

\bibitem[{Shah(2023)}]{Zahir_2023}
Shah, Z. 2023, Monthly Notices of the Royal Astronomical Society, 527, 5140,
  \dodoi{10.1093/mnras/stad3534}

\bibitem[{{Sikora} {et~al.}(1994){Sikora}, {Begelman}, \& {Rees}}]{Sikora_1994}
{Sikora}, M., {Begelman}, M.~C., \& {Rees}, M.~J. 1994, \apj, 421, 153,
  \dodoi{10.1086/173633}

\bibitem[{{Tramacere}(2020)}]{JetSeT_2020}
{Tramacere}, A. 2020, {JetSeT: Numerical modeling and SED fitting tool for
  relativistic jets}, Astrophysics Source Code Library, record ascl:2009.001

\bibitem[{{Tramacere} {et~al.}(2009){Tramacere}, {Giommi}, {Perri},
  {Verrecchia}, \& {Tosti}}]{Tramacere_2009}
{Tramacere}, A., {Giommi}, P., {Perri}, M., {Verrecchia}, F., \& {Tosti}, G.
  2009, \aap, 501, 879, \dodoi{10.1051/0004-6361/200810865}

\bibitem[{Tramacere {et~al.}(2011)Tramacere, Massaro, \&
  Taylor}]{Tramacere_2011}
Tramacere, A., Massaro, E., \& Taylor, A.~M. 2011, The Astrophysical Journal,
  739, 66, \dodoi{10.1088/0004-637X/739/2/66}

\bibitem[{Uemura {et~al.}(2017)Uemura, Itoh, Liodakis, Blinov, Nakayama, Xu,
  Sawada, Wu, \& Fujishiro}]{Ueura_2017}
Uemura, M., Itoh, R., Liodakis, I., {et~al.} 2017, Publications of the
  Astronomical Society of Japan, 69, 96, \dodoi{10.1093/pasj/psx111}

\bibitem[{{Urry} \& {Mushotzky}(1982)}]{Urry_1982}
{Urry}, C.~M., \& {Mushotzky}, R.~F. 1982, \apj, 253, 38,
  \dodoi{10.1086/159607}

\bibitem[{Urry \& Padovani(1995)}]{Urry_1995}
Urry, C.~M., \& Padovani, P. 1995, Publications of the Astronomical Society of
  the Pacific, 107, 803, \dodoi{10.1086/133630}

\bibitem[{Weaver {et~al.}(2020)Weaver, Williamson, Jorstad, Marscher, Larionov,
  Raiteri, Villata, Acosta-Pulido, Bachev, Baida, Balonek, Benítez, Borman,
  Bozhilov, Carnerero, Carosati, Chen, Damljanovic, Dhiman, Dougherty,
  Ehgamberdiev, Grishina, Gupta, Hart, Hiriart, Hsiao, Ibryamov, Joner,
  Kimeridze, Kopatskaya, Kurtanidze, Kurtanidze, Larionova, Matsumoto,
  Matsumura, Minev, Mirzaqulov, Morozova, Nikiforova, Nikolashvili, Ovcharov,
  Rizzi, Sadun, Savchenko, Semkov, Slater, Smith, Stojanovic, Strigachev,
  Troitskaya, Troitsky, Tsai, Vince, Valcheva, Vasilyev, Zaharieva, \&
  Zhovtan}]{Weaver_2020}
Weaver, Z.~R., Williamson, K.~E., Jorstad, S.~G., {et~al.} 2020, The
  Astrophysical Journal, 900, 137, \dodoi{10.3847/1538-4357/aba693}

\bibitem[{Wehrle {et~al.}(2016)Wehrle, Grupe, Jorstad, Marscher, Gurwell,
  Baloković, Hovatta, Madejski, Harrison, \& Stern}]{Wehrle_2016}
Wehrle, A.~E., Grupe, D., Jorstad, S.~G., {et~al.} 2016, The Astrophysical
  Journal, 816, 53, \dodoi{10.3847/0004-637X/816/2/53}

\bibitem[{Weisskopf {et~al.}(2022)Weisskopf, Soffitta, Baldini, Ramsey, O'Dell,
  Romani, Matt, Deininger, Baumgartner, Bellazzini, Costa, Kolodziejczak,
  Latronico, Marshall, Muleri, Bongiorno, Tennant, Bucciantini, Dovciak, Marin,
  Marscher, Poutanen, Slane, Turolla, Kalinowski, Marco, Fabiani, Minuti,
  Monaca, Pinchera, Rankin, Sgr{\`o}, Trois, Xie, Alexander, Allen, Amici,
  Andersen, Antonelli, Antoniak, Attin{\'a}, Barbanera, Bachetti, Baggett,
  Bladt, Brez, Bonino, Boree, Borotto, Breeding, Brienza, Bygott, Caporale,
  Cardelli, Carpentiero, Castellano, Castronuovo, Cavalli, Cavazzuti, Ceccanti,
  Centrone, Citraro, D'Amico, D'Alba, Gesu, Monte, Dietz, Lalla, Persio, Dolan,
  Donnarumma, Evangelista, Ferrant, Ferrazzoli, Ferrie, Footdale, Forsyth,
  Foster, Garelick, Gunji, Gurnee, Head, Hibbard, Johnson, Kelly, Kilaru,
  Lefevre, Roy, Loffredo, Lorenzi, Lucchesi, Maddox, Magazzu, Maldera,
  Manfreda, Mangraviti, Marengo, Marrocchesi, Massaro, Mauger, McCracken,
  McEachen, Mize, Mereu, Mitchell, Mitsuishi, Morbidini, Mosti, Nasimi, Negri,
  Negro, Nguyen, Nitschke, Nuti, Onizuka, Oppedisano, Orsini, Osborne, Pacheco,
  Paggi, Painter, Pavelitz, Pentz, Piazzolla, Perri, Pesce-Rollins, Peterson,
  Pilia, Profeti, Puccetti, Ranganathan, Ratheesh, Reedy, Root, Rubini,
  Ruswick, Sanchez, Sarra, Santoli, Scalise, Sciortino, Schroeder, Seek,
  Sosdian, Spandre, Speegle, Tamagawa, Tardiola, Tobia, Thomas, Valerie,
  Vimercati, Walden, Weddendorf, Wedmore, Welch, Zanetti, \&
  Zanetti}]{IXPE_2022}
Weisskopf, M.~C., Soffitta, P., Baldini, L., {et~al.} 2022, Journal of
  Astronomical Telescopes, Instruments, and Systems, 8, 026002,
  \dodoi{10.1117/1.JATIS.8.2.026002}

\end{thebibliography}
\bibliographystyle{aasjournal}



\end{document}